\documentclass[fleqn,usenatbib]{mnras}

\usepackage{newtxtext,newtxmath}

\usepackage[T1]{fontenc}
\usepackage{ae,aecompl}

\input psfig.tex
\usepackage[english]{babel}
\usepackage{graphicx}
\usepackage{caption}
\usepackage{morefloats}
\usepackage{natbib}
\bibpunct{(}{)}{;}{a}{}{,}
\usepackage{array}
\usepackage{graphics}
\usepackage{latexsym}
\usepackage{amssymb}
\usepackage{amsmath}
\usepackage{fancyhdr}
\usepackage{float}
\usepackage{multirow}
\usepackage{longtable}
\usepackage{lscape}
\usepackage{morefloats}
\usepackage{slashbox}

\usepackage{graphicx}	
\usepackage{amsmath}	
\usepackage{amssymb}	

\title[Red \& Dead CANDELS]{Red \& Dead CANDELS: massive passive galaxies at the dawn of the Universe}

\author[E. Merlin et al.]{E. Merlin$^{1}$,
F. Fortuni$^{1}$,
M. Torelli$^{1}$,
P. Santini$^{1}$, 
M. Castellano$^{1}$, 
A. Fontana$^{1}$, \newauthor
A. Grazian$^{2}$, 
L. Pentericci$^{1}$, 
S. Pilo$^{1}$ and K. B. Schmidt$^{3}$
\\
\\
$^{1}$INAF - Osservatorio Astronomico di Roma, via Frascati 33, 00078 Monte Porzio Catone (RM), Italy\\
$^{2}$INAF - Osservatorio Astronomico di Padova, Vicolo Osservatorio 5, 35122, Padova, Italy\\
$^{3}$Leibniz-Institut f{\"u}r Astrophysik Potsdam (AIP), An der Sternwarte 16, 14482 Potsdam, Germany
}

\date{Accepted 2019 September 11. Received 2019 August 26; in original form 2019 May 22.}

\pubyear{2019}
\begin{document}
\label{firstpage}
\pagerange{\pageref{firstpage}--\pageref{lastpage}}
\maketitle

\begin{abstract}

We search the five CANDELS fields (COSMOS, EGS, GOODS-North, GOODS-South and UDS) for passively evolving a.k.a. ``red and dead'' massive galaxies in the first 2 Gyr after the Big Bang, integrating and updating the work on GOODS-South presented in a previous paper. We perform SED-fitting on photometric data, with top-hat star-formation histories to model an early and abrupt quenching, and using a probabilistic approach to select only robust candidates. Using libraries without (with) spectral lines emission, starting from a total of more than 20,000 $z>3$ sources we end up with 102 (40) candidates, including one at $z=6.7$. This implies a minimal number density of $1.73 \pm 0.17 \times10^{-5}$ ($6.69 \pm 1.08 \times10^{-6}$) Mpc$^{-3}$ for $3<z<5$; applying a correction factor to account for incompleteness yields $2.30 \pm 0.20 \times10^{-5}$. We compare these values with those from five recent hydrodynamical cosmological simulations, finding a reasonable agreement at $z<4$; tensions arise at earlier epochs. Finally, we use the star-formation histories from the best-fit models to estimate the contribution of the high-redshift passive galaxies to the global Star Formation Rate Density during their phase of activity, finding that they account for $\sim$5-10\% of the total star formation at $3<z<8$, despite being only $\sim0.5\%$ of the total in number. The resulting picture is that early and strong star formation activity, building massive galaxies on short timescales and followed by a quick and abrupt quenching, is a rare but crucial phenomenon in the early Universe: the evolution of the cosmos must be heavily influenced by the short but powerful activity of these pristine monsters.

\end{abstract}




\section{Introduction} \label{intro}

Quantifying the abundance of passively evolving (``red and dead'') galaxies in the early Universe is a difficult but crucial task. We know that massive galaxies typically have red colors at all epochs: while in the local Universe this is mostly caused by the absence of young stellar populations \citep[with a degeneracy caused by metallicity, e.g.][]{Worthey1994}, at high redshift this is more often a consequence of high star-formation rates (SFRs) coupled with strong dust obscuration \citep[e.g.][]{Cimatti2002, Dunlop2007}. However, it is established that a non-negligible fraction of massive galaxies in the first $\sim2$ Gyr after the Big Bang is intrinsically red because of passive evolution following the quenching of the star-formation (SF) activity \citep[e.g.][]{Labbe2005,Mobasher2005,Fontana2009,Grazian2015}. 

The very existence of such early red and dead massive galaxies is a challenge to our present understanding of the cosmos. The formation of the structures in the concordance $\Lambda$-CDM cosmological scenario is inherently hierarchical \citep{Press1974, Lacey1993}, with large structures assembling at later times with ongoing bursts of star formation \citep[e.g.][]{White1978, DeLucia2007}. On the other hand, the so-called \textit{downsizing} trend is a well-established evidence, with massive galaxies assembling their stellar content earlier, and typically on shorter timescales, than smaller ones \citep{Matteucci1994, Cowie1996, Thomas2005, Bundy2006, Cimatti2006}. In the last decades, theoretical models and hydrodynamical simulations have struggled to reproduce the properties of the observed galactic populations at all epochs \citep[e.g.][]{Vogelsberger2014, Feldmann2017}; however, to date there is no consensus yet on a robust theoretical approach capable to accurately reconcile the observational data with the models. 

The issue can also be viewed under a different perspective. In the last 15 years the tight correlation between galaxies SFRs and stellar masses, the so-called ``main sequence'', has become a thoroughly studied topic \citep{Brinchmann2004, Noeske2007b, Elbaz2007, Daddi2007}. The sequence is now confirmed to exist to high redshifts \citep{Rodighiero2014, Schreiber2015}. However, at any epoch some (typically compact, bulge-dominated) galaxies fall below it, indicating little or no star formation activity and implying the occurred action of some quenching mechanism \citep{Wuyts2011,Tacchella2018}. The time-scales of such processes, and the physical drivers behind them, remain largely unclear to date \citep[see e.g.][]{Man2018}; the usual suspects include AGN-driven outflows \citep{Brennan2017}, stellar feedback \citep{Kawata1999, Chiosi2002, Ceverino2009, Merlin2012}, gas strangulation \citep{Peng2015} or starvation \citep{Feldmann2015}, virial shocking of the circum-galactic medium \citep{Dekel2006}, or a combination of all these. Whatever the cause, an abrupt halt of the star formation activity makes the galaxy colors turn redder, but blue light from young stellar object can outshine the old populations for several Myrs after the quenching. This makes a simple color-based selection prone to bias, even when using rest-frame inferred magnitudes as in the $UVJ$ diagram \citep{Labbe2005,Wuyts2007}. Galaxies that have quenched shortly before being observed will not enter the selection regions until later times unless an ad-hoc modeling is adopted, and as we have shown in our previous paper \citep[][M18 hereafter]{Merlin2018} in which we exploited the CANDELS  photometric data for the GOODS-South field \citep{Grogin2011,Koekemoer2011,Guo2013}, this is particularly true at very high redshifts, when the timescales of the events are comparable to the life span of the Universe. 
\citet{Davidzon2017} and \citet{Ichikawa2017} argue that the $NUVrJ$ diagram is better suited to identify recently quenched galaxies. However, the CANDELS catalogue does not include a $NUV$ band (0.23 $\mu$m), so we could not counter-check the reliability of such technique. Of course, other more refined approaches - e.g. the analysis of the the main sequence where recently quenched objects are found in a transient position between star forming and passive objects - could be investigated, but the analysis would be in any case posterior to the SED-fitting, since the knowledge of the physical properties of the sources would be required.

In M18 we also showed that tailoring a reliable method to identify high-redshift passive objects is arduous anyway, because of the low signal-to-noise ratio (SNR) of such distant sources, even of the bright ones. This makes it challenging to compare the observed fluxes with template models of Spectral Energy Distributions (SEDs) or colors. In that work, we took advantage of the photometric data from the GOODS-South catalogue, complemented with new observations \citep{Fontana2014} and new deep \textit{Spitzer} mosaics; we exploited state-of-the-art techniques \citep[\textsc{t-phot},][]{Merlin2015,Merlin2016} and ad-hoc SED-fitting libraries built with constant (a.k.a. ``top-hat'') star formation histories (SFHs); and we used a stringent statistical approach to exclude potential false positives. In this way we ended up with 30 passive candidates at $z>3$. However, we also showed how changing some properties of the stellar libraries, or letting the redshifts of the solutions vary, dramatically impacted the results, reducing the sample to 10 (including nebular line emission in the models) or even only two (letting the redshifts free in the fitting process) candidates. To strengthen the robustness of our selection and validate the basic assumptions, in \citet[][ S19 hereafter]{Santini2019} we further checked the nature of the M18 candidates by means of archival ALMA data (available for 26 out of the 30 sources), statistically corroborating the passive classification of the sample from the lack of on-going star formation as seen at sub-mm wavelengths, free from the parameter degeneracies (especially the age-dust degeneracy) typical of the optical domain. Moreover, we could robustly confirm the individual passive nature of 35\% of our candidates, adopting conservative assumptions. In M18 we also showed that upcoming facilities such as the \textit{James Webb Space Telescope} will propel a leap forward, allowing for a much more robust photometric precision and, consequently, determination of physical properties. However, for now we can only trust the predictive and analytic power of the currently available instrumentation, and enlarge the statistical significance including more data. 

To this aim, in the present paper we discuss the results from the joint analysis of the remaining four CANDELS fields (COSMOS, EGS, GOODS-North, and UDS). Since we used a refined grid of SED models, we also repeat the processing on GOODS-South. The five fields have different typical depths, therefore mixing the analysis might be risky, but we can safely consider the results as a lower limit to the actual number of passive objects above $z>3$. Furthermore, in this work we address two more points: the concordance of the observations with the predictions from numerical models, and the impact that these early monsters had on the global SFH of the Universe.



The paper is organized as follows. In Section \ref{method} we describe the dataset and we briefly summarize the method we used to single out the passive sample. In Section \ref{fir} we discuss the confirmation of the candidates by means of the available far-infrared (FIR) and spectroscopic data,  
and in Section \ref{results} we discuss some properties of our candidate galaxies. In Section \ref{ndens} we compute the number densities of our passive sample, and we compare our findings to the predictions of five 
state-of-the-art hydrodynamical models: \textsc{Illustris} \citep{Vogelsberger2014}, \textsc{Illustris}-TNG100 and TNG300 \citep{Pillepich2018}, \textsc{Eagle} \citep{Schaye2015} and \textsc{Simba} \citep{Dave2019}. 
In Section \ref{sfrd} we present a method to compute the Star Formation Rate Density (SFRD) from the fitted SEDs, and compare the contribution of the red and dead populations to the total. Finally, in Section \ref{conclusions} we summarize and discuss the main findings of the work.

Throughout the paper, we assume a $\Lambda$-CDM cosmology ($H_0=70.0$, $\Omega_{\Lambda}=0.7$, $\Omega_m=0.3$), a \citet{Salpeter1959} Initial Mass Function (IMF) except where noted otherwise, and AB magnitudes.

\section{Dataset and methods} \label{method}

For the GOODS-South field we use again the 19-bands catalog already discussed in M18, which improves on the original catalog published by \citet{Guo2013} as it includes three more bands \citep[WFC3 $F$140$W$ from the \textit{Hubble Space Telescope} and VIMOS $B$, plus the deep HAWK-I $Ks$ band presented in][]{Fontana2014}, and it has improved photometry on the \textit{Spitzer} bands thanks to new mosaics (IRAC CH1 and CH2, by R. McLure) and new software (all four channels were re-processed using \textsc{t-phot}). As anticipated we decided to re-analyze the GOODS-South field taking advantage of refined SED libraries and redshift estimates (see below).

For the remaining four fields, we exploited the published CANDELS photometric catalogs, released in 2015 and presented in \citet{Nayyeri2017}, \citet{Stefanon2017}, \citet{Barro2019} and \citet{Galametz2013} for COSMOS, EGS, GOODS-North and UDS respectively.
All catalogs are based on $\sim 20$ wide bands, and in COSMOS and EGS they are complemented with some narrow and/or medium bands. Fluxes have been typically measured by means of \textsc{SExtractor} \citep{Bertin1996} aperture photometry for \textit{Hubble} bands, after PSF-matching to the detection band $H160$; and template-fitting for ground-based and \textit{Spitzer} bands, with \textsc{TFIT} \citep{Laidler2007} or \textsc{t-phot} \citep{Merlin2015, Merlin2016}\footnote{In template-fitting tecnhiques, cutouts from the high-resolution detection band are used as priors to build low-resolution templates of the sources, by means of a convolution kernel that matches the PSFs of the two images. The templates are then used to solve a linear system minimizing the difference between a model collage and the real low-resoluion image, assigning to each source a multiplicative factor that best matches the observed flux. The method has proven to yield great improvements especially when the blending of the sources becomes important, as it is the case for ground based and mid/far-infrared bands. \textsc{t-phot} is the heir of \textsc{TFIT}; it improves on it in terms of accuracy, robustness and computational performance, and it includes a number of additional options. For a detailed description of the techniques and the codes, plase refer to the cited publications.}. The typical depth of the detection band, WFC3 $F$160$W$, is $\sim27.5$ (5$\sigma$ in 2 FWHM diameter). The properties of the five fields are summarized in Table \ref{phottab}; the cumulative area is $\sim$969.7 sq. arcmin.

As for the redshifts, we took advantage of the latest CANDELS estimates, to be presented in Kodra et al. (in preparation) which improve upon the original \citet{Dahlen2013} estimates; the new photo-$z$'s (where spec-$z$'s are not available) are obtained combining four independent estimations, using the minimum Frechet distance combination method. 

In COSMOS, ID-16676 corresponds to the Z-FOURGE source 20115, a recognized passive source first discussed in \citet{Glazebrook2017}, and object of a thorough study by \citet{Schreiber2018a}, who showed how the presence of sub-mm flux is actually due to a strongly obscured close companion. \citet{Glazebrook2017} assigns to this source a spectroscopic redshift of 3.7172, different from both the CANDELS and the 3D-HST photo-$z$'s (4.127 and 3.545, respectively). We take their spec-$z$ as the reference redshift of the object and use it in all our subsequent analysis. In other cases of spectroscopically confirmed redshift we kept the CANDELS photo-$z$ since the estimate was always sufficiently close (see Section \ref{zspec}).





\begin{table}
\renewcommand{\arraystretch}{1.5}
\caption{Summary of the five CANDELS catalogs: number of photometric bands, 5$\sigma$ limiting magnitudes in WFC3 F160W (the detection $H$ band), area in sq. arcmin. 
In COSMOS, 20 out of 43 bands are medium/narrow bands from SuprimeCam and NEWFIRM; in EGS, 6 out of 23 bands are medium bands from NEWFIRM.
In GOODS-North the estimate of the limiting magnitude corresponds to the Wide area; corresponding limiting magnitudes for the Intermediate and Deep areas are 28.2 and 28.7, respectively. A similar pattern applies to GOODS-South, where the estimate of the limiting magnitude corresponds to the Wide area; corresponding limiting magnitudes for the Deep and HUDF areas are 28.16 and 29.74, respectively. 
$m_{\textrm{lim}}$ are typically defined as 5 $\times$ the average standard deviation within 2 FWHM circular apertures in empty regions of the fields.}
\centering
\begin{tabular}{ | l || l | l | l |}
\hline
Field & Bands & $H_{\textrm{lim}}$ & Area \\ \hline\hline
COSMOS & 43 (20) & 27.56 & 216.0 \\ \hline
EGS & 23 (6) & 27.6 & 206.0 \\ \hline
GOODS-N & 18 & 27.8 &  173.0 \\ \hline
GOODS-S & 19 & 27.36 & 173.0 \\ \hline
UDS & 19 & 27.45 & 201.7 \\ \hline
\end{tabular} \label{phottab}
\end{table} 

\subsection{Selection technique and results} \label{technique}

As we did in M18, we proceed as follows to single out our red and dead candidates:
\begin{itemize}
\item we build a library of SED templates with top-hat SFHs, to better model the abrupt quenching of star formation in the very early Universe. The rationale for this choice is extensively discussed in M18.  We improved upon the library used in the previous work by extensively refining the grid of models; the new library is described in detail in Appendix \ref{THlib};
\item we consider all the objects from the $H$-detected catalogs of the five CANDELS fields at $z_{\textrm{CANDELS}} \geq 3.0$, and we select the objects with $H < 27$ and with SNR $ > 1$ in $Ks$, IRAC-CH1 and IRAC-CH2;
\item on these lists of sources, we perform SED-fitting using our code \textsc{zphot} \citep{Fontana2000}, in two flavors: (i) without the inclusion of nebular emission lines (we dub resulting selection ``reference'' sample), and (ii) with the inclusion of the lines (we dub this selection ``lines'' sample). We also made a third run, (iii) including the lines and letting the redshift of the fit vary (``$z$-free'' sample), as in M18. However, S19 have shown that the third criterion is too conservative, given that in M18 only 2 over 30 galaxies survived this selection for the GOODS-South field, while 9 out of 26 sources have been robustly confirmed as passive by means of the follow-up analysis on ALMA data. Therefore, we only cite it here for the sake of completeness, but we will not discuss it further in the paper;
\item we include in our lists of passive candidates only the objects having: (i) a passive best-fit model with probability $p_{\textrm{best}} > 30\%$; and (ii) only star-forming solutions (if they have any) with probability $p_{\textrm{SF}}<5\%$ (these figures come from tailored simulations which are discussed in M18).
\end{itemize}

\begin{figure}
\centering
\includegraphics[width=8cm]{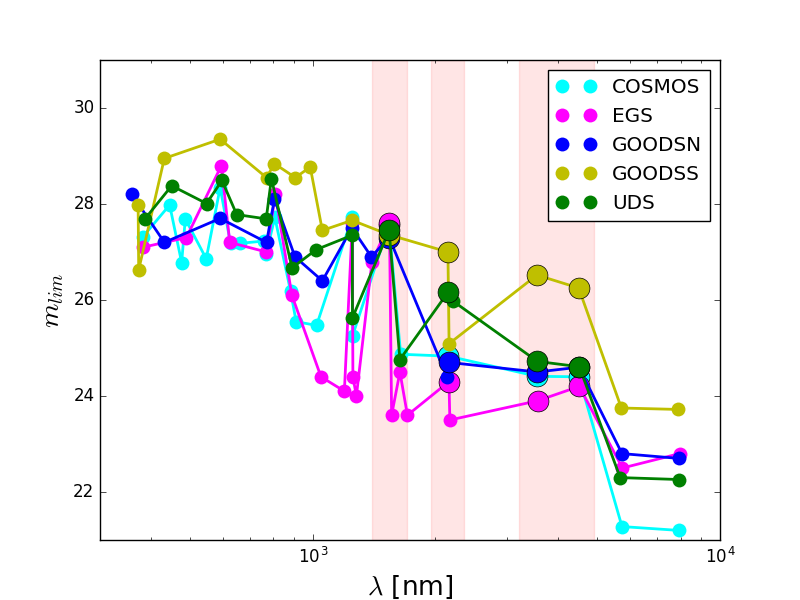}
\caption{Compared $5\sigma$ limiting magnitudes for all the wide bands in the five fields, as given in the CANDELS papers cited in the text. Highlighted with vertical red strips and larger symbols are the bands that are crucial in our analysis, i.e. from left to right WFC3 $H$160 (the detection band, which has similar depths in all the fields), $Ks$, and the first two IRAC channels.} \label{depths}
\end{figure}

\begin{figure*}
\centering
\includegraphics[width=15cm]{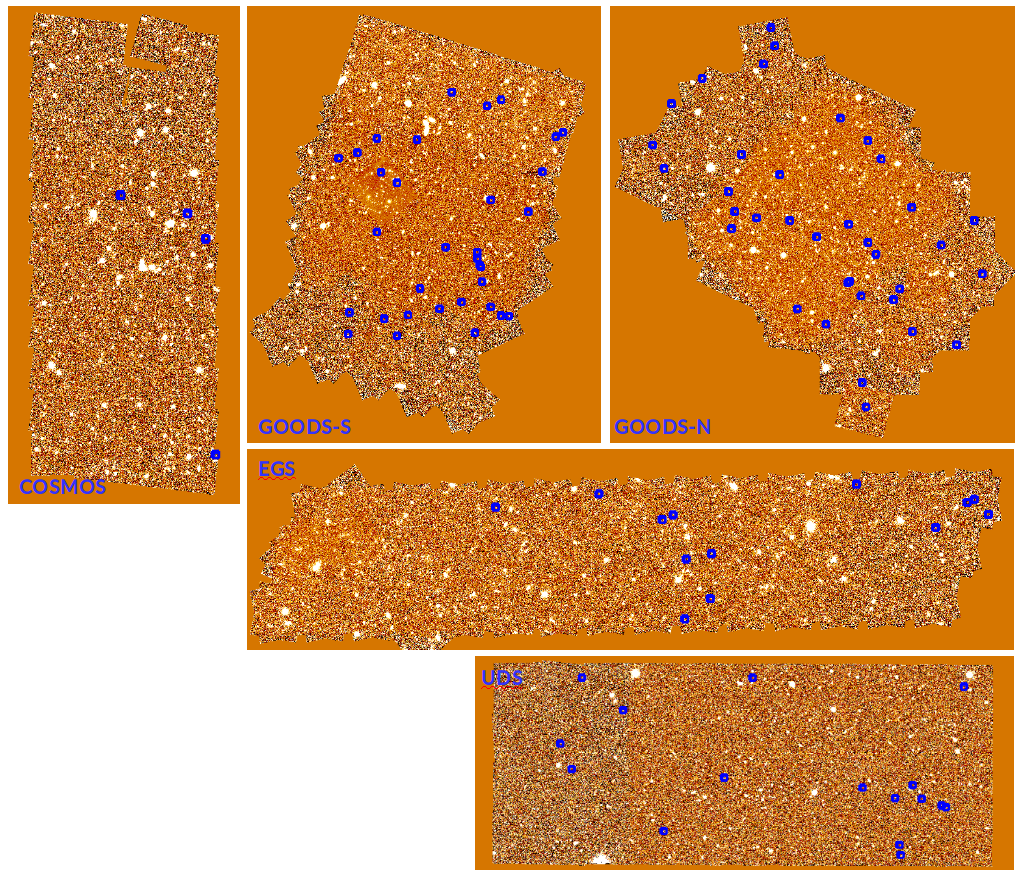}
\caption{Positions of the passive candidates in the five CANDELS fields (shown is the detection band, WFC3 $H160$). The dimensions of the images are not scaled to the real dimensions of the fields. We checked that the sources close to borders have complete image covering. Snapshots of all the candidate galaxies are given in the on-line supplementary material.} \label{allcandidates}
\end{figure*}

In this way we end up with the three selections summarized in Table \ref{selects}. In total, we find 102 candidates in the ``reference'' sample, which become 40 in the ``lines'' selection.
In Fig. \ref{allcandidates} we show the positions of the ``reference'' candidates on the five CANDELS fields.

We point out that while the properties of the detection band ($H$160) are quite similar in the five fields, the same does not hold for what concerns the other bands. In Fig. \ref{depths} the nominal $5\sigma$ limiting magnitudes across the spectrum, as reported in the papers describing the catalogues, are compared: they show that the quality is far from uniform. This is particularly true for the $K$ and \textit{Spitzer} bands, which are crucial for the characterization of high-$z$ objects:
\begin{itemize}
    \item the $K$ bands come from various surveys and facilities, so that their properties are very different in the five fields. For example, the HAWK-I $Ks$ mosaic in GOODS-South is very deep ($m_{\textrm{lim}}\simeq27.0$, 5$\sigma$ in 2 FWHM) with the finest seeing (FWHM$\simeq$0.4''), ensuring exquisite quality data. On the other hand, UDS HAWK-I $Ks$ has FWHM$\simeq$0.4'' and $m_{\textrm{lim}}\simeq25.9$, the COSMOS VISTA $Ks$ band has FWHM$\simeq$0.98'' and $m_{\textrm{lim}}\simeq24.8$, EGS WIRCAM has FWHM$\simeq$0.65'' and $m_{\textrm{lim}} \simeq 24.3$, and GOODS-North CHFT WIRCam has FWHM$\simeq$0.6'' and $m_{\textrm{lim}}\simeq24.7$. The filter response functions are different as well;
    \item the IRAC bands also reach very different depths in the various fields: for example, CH1 reaches $m_{\textrm{lim}}=26.5$ at 5$\sigma$ in GOODS-S, while it is limited to 24.7 in UDS,  24.5 in GOODS-N, 24.4 in COSMOS, and 23.9 in EGS. 
\end{itemize}

Of course, these differences have a strong impact on the efficiency of our methods in the five fields, resulting in significant variations of the number of candidates, as Tab. \ref{selects} shows.

It is worth pointing out that the new redshift estimates change the selection for the GOODS-South field with respect to our sample in M18. Of the present 33 candidates, 26 are in common with the previous selection; 4 are now excluded (IDs 5592, 9091, 26802, 10759) and 7 new are included (IDs 3718, 4202, 4949, 5934, 13394, 16526, 19883). All of the changes are due to variations in the probabilities of star forming solutions in the SED-fitting procedure caused by the different photo-$z$. Since we checked that our candidates were reliable both in M18 and in S19, this seem to imply that our selection criteria are conservative, and the true number of passive objects is probably higher than these estimates.

We also note that IDs 2075 in COSMOS; 24177 in EGS; 157, 643, 6620, 9626, 13007, 24572 and 27251 in GOODS-North; 4949, 6407, 12178 and 19446 in GOODS-South, show complex morphologies in the optical, $H$ or $K$ bands, and possibly have close companions which might cause strong contamination despite the robustness of the adopted photometric methods.

\begin{table*}
\renewcommand{\arraystretch}{1.5}
\caption{Number of sources in the five CANDELS fields, and in total. Left to right, we list: the total number of detected sources, the number of sources at $z_{\textrm{CANDELS}}>3$, the number of sources at $z_{\textrm{CANDELS}}>3$ with SNR>1 in $K$, IRAC1 and IRAC2; and the number of selected candidates in the two passive selections: the ``reference'' sample and the ``lines'' sample obtained including nebular lines in the library.} 
\centering
\begin{tabular}{ | l || l | l | l | l | l |}
\hline
Field/Sample & Total & $z>3$ & S/N$_{z>3}>1$ & Reference & Lines  \\ \hline\hline 
COSMOS & 38671 & 3778 & 1525 & 4 & 2 \\ \hline 
EGS & 41457 & 4830 & 1775 & 13 & 5 \\ \hline 
GOODS-N & 35445 & 3953 & 1793 & 36 & 11 \\ \hline 
GOODS-S & 34930 & 5029 & 2884 & 33 & 13 \\ \hline 
UDS & 35932 & 4018 & 2540 & 16 & 9 \\ \hline 
All fields & 186435 & 21608 & 10517 & 102 & 40 \\ \hline 
\end{tabular} \label{selects}
\end{table*}

\section{Confirmation of candidates with far-infrared and spectroscopic data} \label{fir}

In this Section we study the properties of our selected candidates using their far-infrared photometry and spectroscopic data, when available.

\subsection{\textit{Herschel} fluxes}

As in M18, we have cross-matched the positions of our selected candidates with \textit{Herschel} public catalogs, to check for degeneracies and possible misinterpretations of low-$z$ dusty galaxies as high-$z$ passive dust-free sources. FIR data is available for all CANDELS fields. 

For the two GOODS fields we take advantage of the new, deep \textsc{Astrodeep} catalogs by Wang et al. (in preparation), which combine data from PEP \citep{Lutz2011} and GOODS-Herschel \citep{Elbaz2011} surveys with the PACS camera, and the HerMES survey \citep{Oliver2012} with SPIRE. The catalogs are extracted using $H$-band CANDELS priors, making the cross-correlation very handy. We find that all the new candidates in GOODS-S have no match, while we refer the reader to M18 for the discussion on the two candidates having a FIR-counterpart (IDs 3973 and 10578). The fact that these two sources, also showing $X$-ray emission, are among the ones robustly confirmed by ALMA (see S19), consolidates their interpretation as galaxies that have been recently quenched by strong feedback from the central active nucleus, the latter likely being the responsible for the detected FIR emission. We find one possible match among GOODS-N candidates (ID-35028), having a 3.5$\sigma$ flux at 100 $\mu$m and a 1.5$\sigma$ flux at 160 $\mu$m. However, its flux may be strongly blended with that of a very close-by source, only 0.8'' apart, with a lower redshift ($z\sim 1.8$, more common for \textit{Herschel} detections) and a low level of on-going SF (0.2~$M_\odot$/yr according to the fit with standard exponentially declining star formation histories, or $\tau$-models).

For the other fields we used the HerMES DR4 (COSMOS and EGS) and DR3 (UDS) catalogs \citep{Roseboom2010,Roseboom2012} for the SPIRE bands, while for PACS bands we used the PEP prior-based catalogs (COSMOS and EGS) and the catalogs compiled by the HerMES team (UDS). 
The only source possibly showing some FIR emission is
ID-2075 in COSMOS: it is associated to a source at a distance of 2'' having a 3$\sigma$ flux at 250 $\mu$m, but the flux estimate could be contaminated by a brighter, star-forming source 6'' apart (according to the fit with $\tau$-models; we remind that the PSF at 250 $\mu$m is 18'').

Given these results, and considering the high fraction of potential mis-associations due to the large \textit{Herschel} PSFs, and the contamination from nearby sources, we can conclude that we did not find any clear evidence for FIR emission for any of our ``reference'' candidates, with only a few moderately uncertain cases.

\subsection{Spectroscopic data} \label{zspec}

Checking the new CANDELS catalogs to be presented in Kodra et al., we find that the following candidates have spectroscopic redshifts (which were used in our analysis): IDs 2490 and 6539 in EGS, IDs 20589 in GOODS-North, IDs 10578 and 16526 in GOODS-South, and ID 8689 in UDS. Among these, only GOODSS-10578 enters our ``lines'' selection, while all the other are only in the ``reference'' sample.

We searched the VANDELS \citep{Pentericci2018, McLure2018} and VUDS \citep{LeFevre2015} databases to visually inspect the available spectra for GOODS-South, finding data for additional sources with respect to the CANDELS catalogue. All the spec-$z$'s are consistent with the CANDELS estimates, unless explicitly specified. 
We already discussed IDs 4503, 9209 and 10578 in M18 (Sect. 4.4). To these, we can now add from the DR2: IDs 4949 and 5934, which show no evident features; ID-12178, which has a strong emission line and is classified at $z_{\textrm{spec}}=0.56$ (while $z_{\textrm{CANDELS}}=3.29$), but might be spectroscopically contaminated by a very close companion (see Appendix \ref{examples}); and IDs 16526 and 19505, both showing moderate Lyman-$\alpha$ emission, but no other evident features.

Many candidates in GOODS-South were also observed with MUSE, as part of the MUSE-Wide \citep{Urrutia2018} and MUSE Deep \citep{Bacon2017, Inami2017} GTO programs (while all our COSMOS candidates all fall outside the MUSE-Wide COSMOS footprint). 
Most of the candidates covered by the 1 hour deep extended MUSE-Wide data (GOODS-South IDs: 3912, 4587, 4949, 5934, 6407, 7526, 7688, 8242, 8785, 9209, 12178, 17749, 18180, 19301, 19446) show no signs of Ly$\alpha$ emission, in support of these sources candidacy as passive galaxies. Although in general the spectra are not deep enough to provide conclusive evidence, the absence of detectable emission lines strengthen our conclusions. For IDs 9091, 12178 (already mentioned above) and 13394 there is line emission detected within 1.0'', 0.5'' and 0.5'' in the MUSE-Wide data, respectively. These emission lines are however associated with different foreground objects at redshifts 1.33, 0.56 and 0.42 (MUSE-Wide DR1 ID: 143003008). Three candidates in GOODS-South fall in the MUSE Deep footprint. ID-15457 has no emission detected in the MUSE Deep data; approximately 0.5'' from ID-16506, a Ly$\alpha$ emitter is detected at $z=3.33$. Lastly, ID-10578 is confirmed as an AGN by the MUSE Deep data, as already discussed in M18.

Based on these results, we conclude that we find no strong spectroscopic evidence to exclude any of our candidates from the selected sample.


\section{Properties of the selected sample} \label{results}

The main physical properties of all the candidate objects, as obtained in the SED-fitting procedure with the top-hat library, are given in Appendix \ref{phystab}. We provide snapshots and SEDs of all the candidates in the on-line supplementary material, and we show a few significant examples in Appendix \ref{examples}. Here we give a brief summary of the global properties of the candidates.

First of all, we compared the properties of our 102 ``reference'' candidates with those from the 3D-HST catalogs \citep{Skelton2014}. Most of the 3D-HST photometric redshifts, obtained with \textsc{EAzY} \citep{Brammer2008}, are in reasonably good agreement with the CANDELS ones ($\Delta z < 0.3$), with the exceptions of the 19 sources listed in Tab. \ref{3d} (inconsistencies between the two catalogs can be due to many factors, including the usage of different photometric methods and photo-$z$ codes). None of the discrepant objects has a spectroscopic redshift estimate; most of them have lower photo-$z$ in 3D-HST than in CANDELS, with 8 candidates that would be excluded from our selections having $z_{\textrm{3D-HST}} < 3$ and one (ID-13 in GOODS-North) lacking a reliable photo-$z$ estimate. 

We note that in GOODS-South 6 out of 9 discrepant objects are among our strongest candidates (i.e. they belong to the ``lines'' selection
), with three among them having been individually confirmed as passive sources by the analysis in S19 on ALMA data: IDs 9209, 17749 and 18180. However, in 3D-HST their photo-$z$ is still above 3, and all have $sSFR<10^{-11}$ yr$^{-1}$ \citep[estimated with \textsc{FAST},][]{Kriek2009}, therefore being robust candidates in their analysis as well. More in general, while $\sim38\%$ of our candidates having $z_{\textrm{3D-HST}}>3$ can be labelled as passive using 3D-HST \textsc{FAST} estimates and the hard threshold $sSFR<10^{-11}$ yr$^{-1}$, in $\sim40\%$ of cases the sSFR is above $10^{-10}$ yr$^{-1}$. We note that in contrast, using CANDELS estimates on SFR and masses, we find  $\sim14\%$ and $\sim47\%$, respectively.

\begin{table}
\renewcommand{\arraystretch}{1.5}
\caption{CANDELS vs. 3D-HST photometric redshift outliers ($\Delta z \geq 0.3$). The symbol $\dag$ highlight the sources that have $z_{\textrm{3D-HST}}<3$.} 
\centering
\begin{tabular}{ | l | l | l | l | l |}
\hline
Field & ID$_{\textrm{CAND}}$ & $z_{\textrm{CAND}}$ & ID$_{\textrm{3D}}$ & $z_{\textrm{3D-HST}}$\\ \hline\hline
COSMOS & 16676 & 4.127 & 19670 & 3.5446 \\ \hline \hline
EGS & 25724 $\dag$ & 3.795 & 33354 & 2.8674 \\ \hline \hline
GOODSN & 13 & 3.014 & 5 & - \\ \hline 
GOODSN & 1570 $\dag$ & 3.226 & 2212 & 0.1497 \\ \hline
GOODSN & 5744 & 3.459 & 8249 & 3.0668 \\ \hline
GOODSN & 10672 $\dag$ & 6.713 & 15083 & 1.7364 \\ \hline
GOODSN & 13403 $\dag$ & 3.793 & 18781 & 0.5635 \\ \hline
GOODSN & 21034 & 3.328 & 28997 & 4.2929 \\ \hline
GOODSN & 28344 & 4.758 & 3484 & 4.3757 \\ \hline
GOODSN & 35028 $\dag$ & 3.642 & 34844 & 1.2655 \\ \hline \hline
GOODSS & 2608 & 3.720 & 4857 & 3.4116 \\ \hline
GOODSS & 3912 & 3.897 & 7177 & 4.9983 \\ \hline
GOODSS & 7526 $\dag$ & 3.317 & 15494 & 2.647 \\ \hline
GOODSS & 8785 $\dag$ & 3.852 & 17894 & 2.5811 \\ \hline
GOODSS & 9209 & 4.486 & 18684 & 4.8638 \\ \hline
GOODSS & 16506 $\dag$ & 3.382 & 30821 & 0.3737  \\ \hline
GOODSS & 17749 & 3.697 & 32872 & 3.1606 \\ \hline
GOODSS & 18180 & 3.650 & 33566 & 3.3366 \\ \hline
GOODSS & 19301 & 3.592 & 35502 & 4.0423 \\ \hline \hline
UDS & 25893 & 4.491 & 41274 & 3.8881  \\ \hline
\end{tabular} \label{3d}
\end{table}

\subsection{UVJ diagram}

\begin{figure}
\centering
\includegraphics[width=9cm]{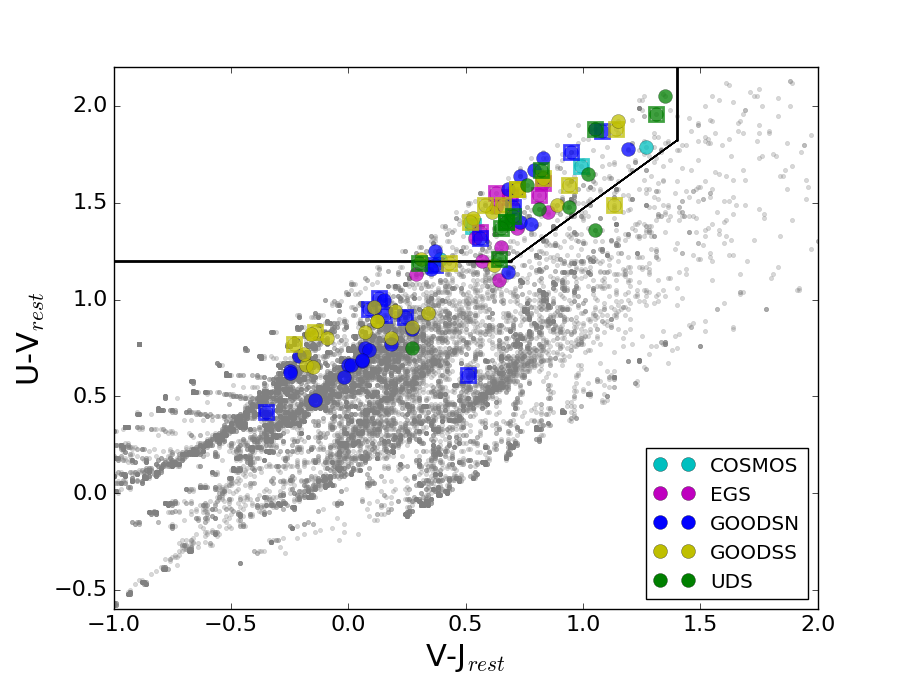}
\caption{UVJ selection diagram. The solid lines enclose the region of passive sources, as identified in \citet{Whitaker2011}. Grey dots: all CANDELS $z>3$ sources. Large coloured dots: passive ``reference'' selection; the dots with a square are also part of the ``lines'' selection. The rest-frame colors are obtained with $\tau$-models fits. The points follow regular patterns because they correspond to the colors of the templates in the SED-fitting library, which have well determined values.} \label{uvj}
\end{figure}

Fig. \ref{uvj} shows the position of our passive candidates on the $UVJ$ diagram. The region of passively evolving objects is delimited as in \citet{Whitaker2011}. Grey small dots correspond to the whole sample of $z>3$ galaxies in the CANDELS fields, while the passive candidates in the ``reference'' sample are plotted as large dots; empty squares mark the candidates belonging to the ``lines'' selection as well.

We point out that for this plot we use the rest-frame colors obtained fitting their observed SEDs with a standard exponentially declining SFH ($\tau$-models): this choice is motivated by the fact that we want to check whether the $UVJ$ color selection method, straightforwardly applied, can be considered a reliable approach. However, in M18 we showed that the colors obtained with our top-hat SFHs are quite similar, with small shifts in the $V-J$ color which we ascribe to the less constrained photometry in the observed redder part of the spectrum (the two 5.6/8.0 $\mu$m IRAC bands have the poorest SNR), with respect to the visible and near-infrared (NIR) bands which straddle the rest-frame $U-V$ break at $z\sim3$. We thus showed how the $UVJ$ selection criteria is certainly powerful, but at these high redshifts it can miss a number of interesting candidates, in particular many  recently quenched objects which still show bluer colors than the typical red passive galaxies.
On the same note, \citet{Schreiber2018b} claimed that the $UVJ$ selection tends to be pure (although with a $\sim20\%$ failure rate) but incomplete, in that a fraction of $\sim40\%$ more quiescent galaxies can be identified using e.g. their specific star-formation rate (sSFR) fitted value.

Here too we see that many candidates lie outside the passive region of the diagram. It is interesting to note that most of the outliers belong to two fields, GOODS-S and GOODS-N, which have high quality data in the infrared bands, in particular considering IRAC CH3 and CH4: this seems reasonable, since their SED-fitting is better constrained, yielding less possible solutions and therefore excluding potential star-forming fits. This seems to indicate that in the other fields we are probably underestimating the real number of passive sources.

More generally, we also note that many objects fall inside the passive region of the diagram, but do not belong to our samples. We ascribe this fact to our stringent criteria, which are tailored to select strictly \emph{passive} objects, rather than \emph{quiescent} ones (we emphasize that we use the term ``quiescent'' to dub sources that retain a weak star-formation activity, as opposed to completely ``passive'' ones). As already noted, it might be too conservative excluding potentially reliable candidates because of the mere existence of a few possible, albeit improbable, star-forming solutions in the SED-fitting procedure.

\subsection{Diagnostic diagrams}

\begin{figure}
\centering
\includegraphics[width=9cm]{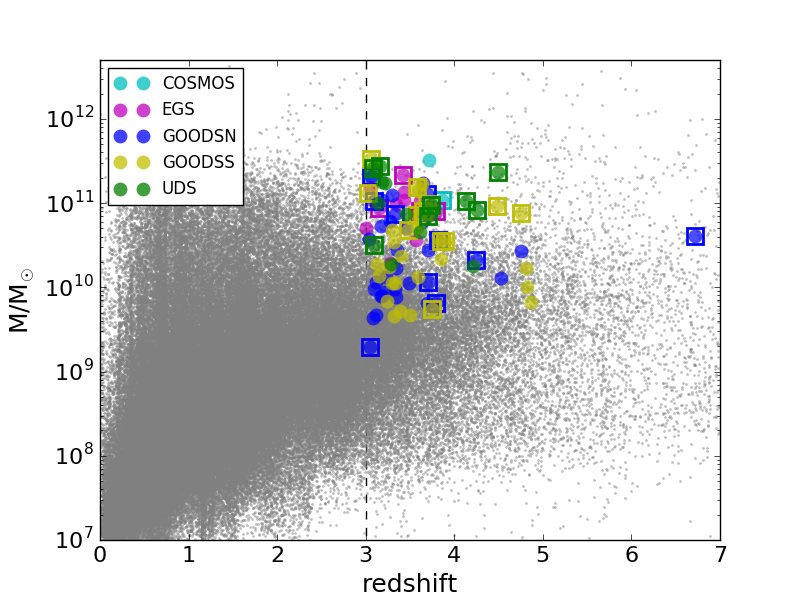}
\caption{Stellar mass vs. redshift of the selected candidates in all CANDELS fields. Grey points show the full CANDELS catalogues, with ``delayed'' $\tau$ SFH mass estimates from \citet{Santini2015}, corrected to account for the Chabrier IMF (our fit assumes a Salpeter IMF). Full circles correspond to the objects from ``reference'' sample; empty squares mark the objects from the ``lines'' selections. For the passive candidates we plot the masses from the best top-hat fit in the ``reference'' run.} \label{zmass}
\end{figure}

Figure \ref{zmass} shows the stellar mass vs. redshift plane. In this case, for the CANDELS full catalog we consider the masses obtained in the best fit with the ``delayed'' $\tau$-models, for which SFR $\propto (t^2/\tau)\times exp(-t/\tau)$ \citep[][]{Santini2015}, while for the red and dead sample we use the ones of the best top-hat fit from this work. We applied a correction factor of 1.75 to make the CANDELS mass estimates, which assume a \citet{Chabrier2003} IMF, consistent with ours, which assume a Salpeter IMF. Most of the ``reference'' candidates have $M \geq 10^{10} M_{\odot}$, which is unsurprising since our selection criteria only select luminous (i.e. high SNR) sources. In particular, we find two candidates at $z>4$ with masses above $10^{11} M_{\odot}$: UDS-10430 and UDS-25893. 

We then consider the half-light radii of $3<z<5$ sources, as measured by the \textsc{SExtractor} detection runs in the $H160$ band, converting them from pixel space to arcseconds and then to proper kpc at the relevant redshifts by means of appropriate \texttt{Astropy} \citep{Astropy2013, Astropy2018} routines.

We do not take into account the $k$-correction factor which would be necessary wince we consider galaxies at different redshifts, implying that the observed $H$ band samples different regions of the rest-frame spectrum (i.e., 400 nm at $z=3$, 320 nm at $z=4$ and 267 nm at z=5). However, we have checked that the sizes measured on images corresponding to bands sampling the same region of the spectrum (i.e., $Y$-band at $z=3$ and $J$-band at $z=4$, at 267 nm rest-frame; all images were PSF-matched to $H$) are very similar to the $H$ band sizes, and considering that the $Y$-band data is not always available we preferred to simply use $H$ band data. A possible drawback of this choice is that in this way we are looking at the rest-frame $B$ band, which is not particularly well suited to study the size of a red object.

The resulting mass vs. radius relation is plotted in Figure \ref{mr}. In general, the passive candidates appear to be typically massive but mostly concentrating towards the region of small radii. Also, a dependence on the redshfit seems to be present, as highlighted by the size-coding of the dots and in the lower panel of the Figure: earlier sources appear to be more compact than later ones. The typical radii of these objects are roughly consistent with the ones found for other high-redshift samples of passive objects, $R_e \sim 1$ kpc, therefore showing a general compactness with respect to local galaxies \citep[e.g.][for $1.4<z<2$ passive galaxies in a similar mass range]{Cimatti2008}. 

\subsection{A massive red and dead galaxy at $z \sim 6.7$?} \label{z7}

We find a particularly interesting object among the galaxies in our selection: GOODS-North ID-10672 has $z_{\textrm{CANDELS}}=6.713$, and it enters both the ``reference'' and ``lines'' samples. 
Its mass inferred from the SED-fitting is $4\times10^{10} M_{\odot}$, with an age of $\sim 500$ Myr, implying a formation redshift of $ \sim 14$, i.e. $ \sim 300$ Myr after the Big Bang; the fitted burst duration is of just 300 Myr, implying an average SFR of $\simeq 130 M_{\odot}$/yr. With a typical $\tau$-model SFH, it is fitted with $sSFR=5.21877\times10^{-11}$ yr$^{-1}$, so it would be classified as quiescent, but it would fail a strictly passive selection based on a standard $sSFR<10^{-11}$ yr$^{-1}$ criterion. The object has been observed and catalogued in many surveys \citep[e.g.][]{Finkelstein2015,Bouwens2015,Harikane2016} as a $z\sim6-7$ mildly star forming or quiescent galaxy. As already pointed out (Tab. \ref{3d}), in 3D-HST it is instead classified a $z=1.73$ source by means of \textsc{EAzY} photometric redshift estimate (without grism information). We find that no detectable signal is associated to this galaxy in MIPS, PACS and SPIRE (i.e. 24 to 500 $\mu$m): while some flux is detected within a few arcsec from the source position, it can quite confidently be attributed to other nearby star-forming objects. However, from a SED-fit with redshift fixed at the 3D-HST value of 1.73 we obtain SFR values of $\sim 3-8 M_{\odot}$/yr (depending on the details of the adopted models), which is too low to be detected in MIPS or \textit{Herschel}. Another possibility might be that the source is actually a cold star, possibly a brown dwarf; instead, we tend to rule out the options of a young and obscured star or of an AGB star, because the source is very isolated, which would be unusual for these kind of objects. A spectroscopic observation would be important to definitively rule out alternative possibilities, but the very faint continuous emission would make the analysis very difficult and perhaps unfruitful. Snapshots and best-fitting SED of this objects are show in Appendix \ref{examples}; a more extended analysis to exclude the possibility of degeneracies in the fitting would be required, but is beyond the scope of the present work and we leave it to future work.

\subsection{Comparison with other selections}

In M18 we compared our results on GOODS-South to the ones by \citet[][S14]{Straatman2014} and \citet[][N14]{Nayyeri2014}. We repeat the analysis now, given that our sample has sligthly changed because of the new redshift estimates. Cross-correlating the CANDELS catalogue with the ZFOURGE selection in S14, we now find that four out of six sources in the S14 selection belong to our sample as well: ID-19883 has now entered both our ``reference'' and ``lines'' selections, together with IDs 4503, 17749 and 18180 which were already present in M18. As for the N14 selection, the situation in unchanged, with five sources in common and the other 11 rejected in our analysis. We refer the reader to M18 for other considerations about these comparisons. 

More recently, \citet{Schreiber2018b} used MOSFIRE $H$ and $K$ spectra to confirm the quiescent nature of 22 galaxies (out of 24 reduced spectra) detected in COSMOS, EGS and UDS, based on ZFOURGE and 3D-HST catalogs. These sources had been previously singled-out as quiescent candidates via $UVJ$ selection. We cross-matched the ZFOURGE, 3D-HST and CANDELS catalogs RA-DEC coordinates, and found that out of the 22 objects, only 9 are in our selection as well (see Tab. \ref{tabbb}). We could not find matching objects for three spectroscopic sources (ZF-COS-17779 is outside the HST footprint, ZF-UDS-35168 and ZF-UDS-39102 are not detected in CANDELS); among the other 9, two have $z_{\textrm{CANDELS}}<3$, while the other 7 do not match our selection criteria (having some SF solutions; we must point out that they select quiescent rather than totally passive sources).

\begin{table}
\renewcommand{\arraystretch}{1.5}
\caption{Matches between the ZFOURGE/3D-HST and CANDELS IDs in the \citet{Schreiber2018b} sample.}
\centering
\begin{tabular}{ | l | l |}
\hline
ZFOURGE / 3D-HST & CANDELS  \\ \hline\hline
ZF-COS-20115 & COS 16676\\ \hline
3D-EGS-18996 & EGS 14727 \\ \hline
3D-EGS-31322 & EGS 24177 \\ \hline
ZF-UDS-3651 & UDS 1244\\ \hline
ZF-UDS-4347 & UDS 2571 \\ \hline
ZF-UDS-7329 & UDS 7520 \\ \hline
ZF-UDS-7542 & UDS 7779 \\ \hline
ZF-UDS-8197 & UDS 8682 \\ \hline
ZF-UDS-41232 & UDS 25688 \\ \hline
\end{tabular} \label{tabbb}
\end{table}

\begin{figure}
\centering
\includegraphics[width=9cm]{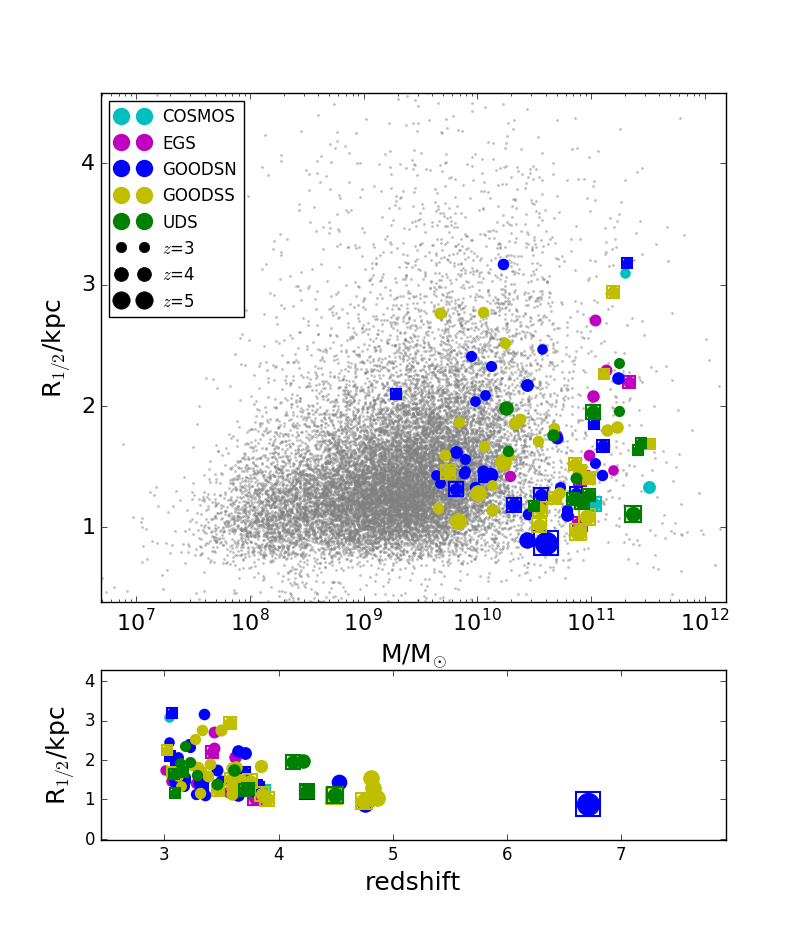}
\caption{Top: mass-radius relation. Grey dots: all CANDELS sources at $3<z<5$; colored circles: passive candidates, size-coded for redshift (see legend); empty squares mark the objects in the ``lines'' selection. The radii are the half-light radius in $H$ band from the \textsc{SExtractor} runs, converted into proper kpc. The masses are the ``delayed'' $\tau$ SFH estimates from \citet{Santini2015}, corrected to account for the Chabrier IMF (our fit assumes a Salpeter IMF). Most of the candidates are on the compact tail of the distribution, with a clear trend with redshift: high redshift sources are smaller compared to $z\sim3$ sources, as show in the bottom panel, where we plot the radii of the candidates against their redshift.}
\label{mr} 
\end{figure}

\section{Number density} \label{ndens}

\begin{figure}
\centering
\includegraphics[width=8.5cm]{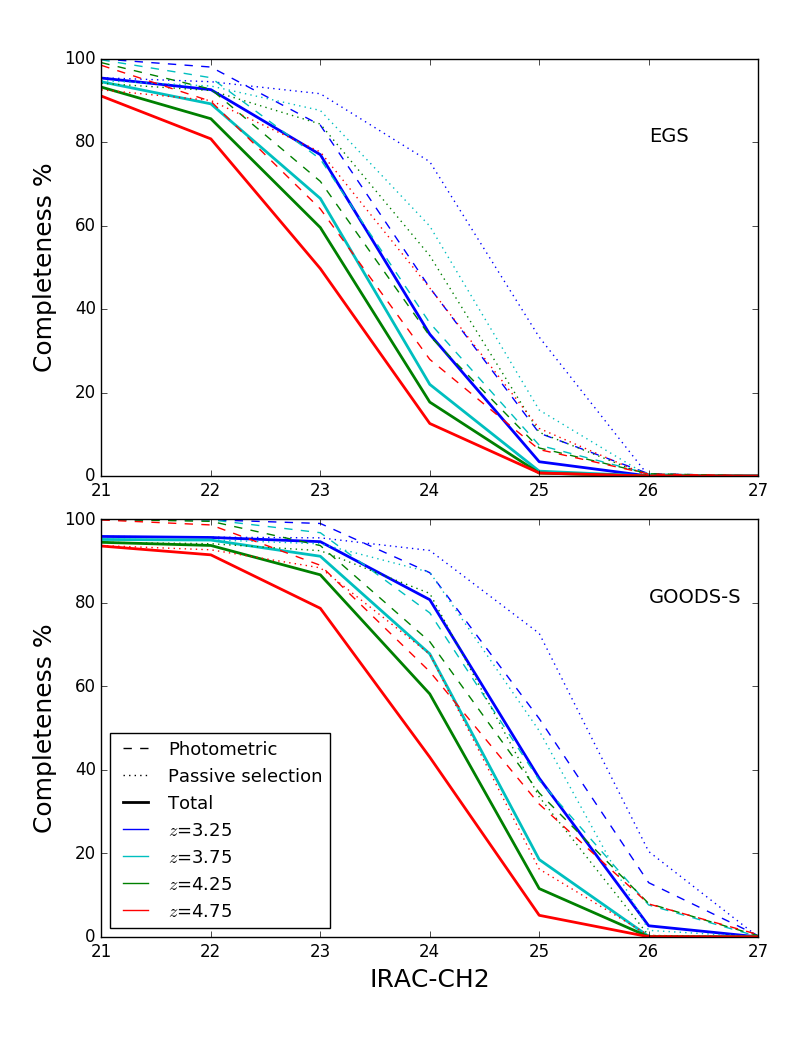}
\caption{Estimated completeness of the passive selection as a function of IRAC-CH2 observed magnitude, as estimated from a set of dedicated simulations, for the two fields with the best and worst data quality (upper panel: GOODS-South; lower panel: EGS). In each panel, the dashed lines correspond to the photometric selection completeness, the dotted lines to the passive candidates selection completeness (computed considering only the photometrically selected galaxies), and the solid lines to the total completeness given by the product of the two. The colors of the lines correspond to different redshift bins. See text for more details.}
\label{compl1} 
\end{figure}

\begin{figure}
\centering
\includegraphics[width=8.5cm]{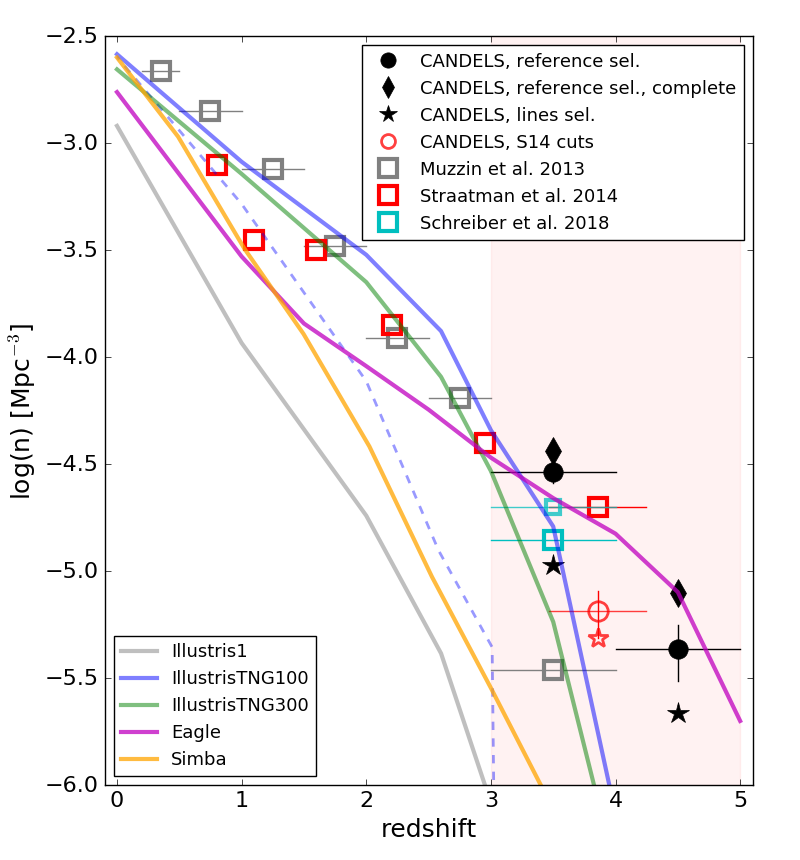}
\caption{Number density of passive galaxies as a function of redshift. We plot many different  estimates, both from observations \citep{Muzzin2013,Straatman2014,Schreiber2018b} and from hydrodynamical simulations (from which we take the galaxies with $M_*/M_{\odot}>5\times10^9$): \textsc{Illustris}-1 \citep{Vogelsberger2014}, \textsc{Illustris}-TNG100 and \textsc{Illustris}-TNG300 \citep{Pillepich2018}, \textsc{Eagle} \citep[][run RefL0100N1504 with full physics included]{Schaye2015}, and \textsc{Simba} \citep{Dave2019}. See text for relevant information on these datasets and models. We then plot the results based on our SED-fitting method for the ``reference'' sample, considering the cuts in mass and redshift adopted by S14 (red empty circle), and the full $3<z<5$ interval divided in two bins considering all masses, before (black circles) or after (black diamonds) correcting for completeness as described in Sect. \ref{completeness}. We also plot the values corresponding to the ``lines'' selections (black or red stars). The shadowed area highlights the relevant redshift interval. The agreement between all observations and the \textsc{Illustris}-TNG models is reasonable up to $z\simeq4$, while it worsens at earlier epochs; on the other hand, \textsc{Eagle} yields better results at $z>4$, but underestimates the number of passive objects at $z<3$. Finally, \textsc{Simba} seem to fall short at all epochs, however doing better than the original \textsc{Illustris}-1 model. See text for more details.}
\label{numdens} 
\end{figure}

We now try to estimate the number density of passive galaxies, dividing the number $N$ of our red and dead candidates observed within a redshift interval of interest $\Delta z $ by the cosmological volume $V$ corresponding to such interval, 
re-scaled by the ratio between the total area of the survey
($\sim 969.7$ sq. arcmin) and that of the full sky. We compute the errors as the poissonian uncertainties $\sqrt{N}/V$, which we assume do take into account cosmic variance given the fact that we are considering five different realizations (fields). As a reference result, we first compute the number densities yielded by our list of candidates, without considering any correction for incompleteness; we will discuss and complement this in the next subsection.

The resulting number density of passive galaxies in the redshift interval $3<z\leq5$, considering the sum of our samples in the five fields, is $1.73 \pm 0.17 \times10^{-5}$ for the ``reference'' selection ($6.69 \pm 1.08 \times10^{-6}$ for the ``lines'' selection) Mpc$^{-3}$. These values would change to $2.16 \pm 0.22 \times10^{-5}$ ($8.52 \pm 1.54 \times10^{-6}$) Mpc$^{-3}$ if we excluded COSMOS from the average, a potentially reasonable choice given that this field yields significantly lower estimates with respect to the other four (which is probably due to cosmic variance, but also to worse photometric properties like e.g. the broad FWHM in $Ks$, see Sect. \ref{technique}). However, in the following we will stick to the estimates obtained averaging on all the five fields.

\begin{table}
\renewcommand{\arraystretch}{1.5}
\caption{Number densities (Mpc$^{-3}$) of passive candidates in the total CANDELS area ($\sim969.7$ sq. arcsec), considering three redshift intervals, for our two selections ``reference'' (both as observed and after correction for completeness) and ``lines''.} 
\centering
\begin{tabular}{ | l || l | l | l |}
\hline
$\Delta z$ & \multicolumn{2}{c|}{Reference} & Lines \\ \hline 
& Observed & Corrected & \\ \hline\hline
$3<z\leq5$ & 1.73$\times10^{-5}$ & 2.30$\times10^{-5}$ &  6.69$\times10^{-6}$ \\ \hline 
$3<z\leq4$ & 2.90$\times10^{-5}$ & 3.66$\times10^{-5}$ & 1.08$\times10^{-5}$ \\ \hline 
$4<z\leq5$ & 4.34$\times10^{-6}$ & 7.94$\times10^{-6}$ &  2.17$\times10^{-6}$ \\ \hline 
\end{tabular}\label{tnd} 
\end{table}

\begin{table}
\renewcommand{\arraystretch}{1.5}
\caption{Number densities (Mpc$^{-3}$) of passive candidates in the five CANDELS fields, considering the cosmological volume $3<z\leq5$, for our two selections ``reference'' (both as observed and after correction for completeness) and ``lines''.} 
\centering
\begin{tabular}{ | l || l | l | l |}
\hline
Field & \multicolumn{2}{c|}{Reference}  & Lines\\ \hline 
& Observed & Corrected & \\ \hline\hline
COSMOS & 3.08$\times10^{-6}$ & 3.50$\times10^{-6}$ & 1.54$\times10^{-6}$ \\ \hline 
EGS & 1.05$\times10^{-5}$ & 1.34$\times10^{-5}$ &  4.04$\times10^{-6}$ \\ \hline 
GOODS-N & 3.36$\times10^{-5}$ & 4.38$\times10^{-5}$ &  9.62$\times10^{-6}$  \\ \hline 
GOODS-S & 3.17$\times10^{-5}$ & 4.44$\times10^{-5}$ &  1.25$\times10^{-5}$  \\ \hline 
UDS & 1.32$\times10^{-5}$ & 1.75$\times10^{-5}$ & 7.42$\times10^{-6}$   \\ \hline 
\end{tabular} \label{tnd2}
\end{table}

\subsection{Completeness} \label{completeness}

It is not easy to try and quantify the completeness of our sample of passive candidates, given the different depths of the five fields, and the subtleties of the technique we adopted. We attempt to do so proceeding as follows. Using \textsc{zphot}, we create a library of 3000 $3<z<5$ synthetic spectra with top-hat SFHs, having different ages and duration of SF bursts so that 1740 are passive and 1260 are star forming. Then, we consider the two fields with the highest and the lowest quality of data in terms of depth, namely GOODS-South and EGS, respectively; by means of the in-house software \textsc{Simulcat}, we use the synthetic models to create two simulated observed catalogues, having the properties of the two fields in terms of filters and depths (i.e., signal-to-noise ratios) at all magnitudes and in all the observed bands, and fixing a reference filter (we choose IRAC-CH2) to predefined magnitude values, i.e. $m_{I2} = 21, 22, 23, 24, 25, 26, 27$. In both cases, each model is replicated ten times with a slightly different noise realization, so each catalogue finally contains 17400 passive objects. Finally, we fit these mock observed catalogs with our top-hat library, and proceed to select passive candidate following exactly the same procedure we used on the real data.

The results for the two fields are show in Fig. \ref{compl1}. In both panels, the solid lines show the inferred completeness in four redshift bins, as a function of the reference magnitude in IRAC-CH2. This completeness is the result of the product of two factors: the photometric completeness, which comes from the pre-selection we perform on the observed catalogue, as described in the second bullet of the list at the beginning of Sect. \ref{technique} (namely: $H<27$, SNR$_{K,IRAC1,IRAC2}>1$, plus an additional condition on the SNR of the detection band to take into account the typical detection cut in the original input catalogs, $SNR_{H}>4$), shown as dashed lines in the Figure; and the passive-selection completeness, which we computed considering only the sources that survived the photometric selection, and is shown with dotted lines. Therefore, the product of the two quantities is self-consistent as a total completeness.

As expected, the values drop as a function of the magnitude and of the redshift. In the case of GOODS-South, a 50\% completeness is reached at $m_{IRAC2}\simeq24.75$ at $3<z<3.5$, and at $m_{IRAC2}\simeq23.75$ at $4.5<z<5$; in EGS, the 50\% completeness is reached at $\sim 1$ magnitude brighter. We note that we are $\sim95\%$ complete in GOODS-South at $m_{IRAC2}\simeq23$ at $z=3$.

We now use these values to estimate an inferred ``true'' number densities of sources (we only do so for the ``reference'' selection, for the sake of simplicity). We multiply the number of objects we actually find in each magnitude and redshift bin by the inverse of the corresponding completeness estimate; as a reasonable approximation of the complex features of the different fields, we use the values that we obtained using GOODS-South for the two GOODS catalogues, and the estimates computed on EGS for the other three. To obtain a fair estimate, we take 50\% as the minimum reliable value, and when the completeness drops below this threshold we continue to use it to compute the actual multiplicative factor. Also, if a bin of magnitude and redshift contains zero objects, its counts will of course remain zero. These two points imply that once again we are being conservative, and the obtained number density can still be underestimated. With this approach, we end up with a corrected total number density of $2.30 \pm 0.20 \times10^{-5}$ for the whole redshift interval $3<z<5$ (for the sake of reference, we point out that if we decided not use the 50\% completeness threshold, and instead we used the full completeness functions as obtained from the simulations, we would get a number density of $3.25 \pm 0.24 \times10^{-5}$). 

In Tab. \ref{tnd} we list the number densities both before and after the completeness correction, for the whole survey area, in three redshift intervals; while in Tab. \ref{tnd2} we give the number densities corresponding to the individual cosmological volumes of the five fields, in the full redshift interval $3<z\leq5$. GOODS-South yields the higher value (by a factor of $\sim2$ with respect to the average of the five fields, in the ``reference'' selection), likely because of a better constrained photometry in the infrared bands leading to more robust SED fits. Cosmic variance also plays a role: the number of all $z>3$ sources varies by a factor of $\sim1.5$ between the five fields (see Tab. \ref{selects}).  

\subsection{Comparison with predictions from numerical simulations} \label{sims}

To understand how the values we have obtained fit in the current theoretical scenario, we have compared the number density of our passive candidates at high redshift with the estimates obtained in five 
recent cosmological hydrodynamical simulations: \textsc{Illustris} \citep{Vogelsberger2014nat,Vogelsberger2014,Genel2014,Nelson2015}, \textsc{Illustris}-TNG100 and TNG300 \citep{Pillepich2018,Nelson2019}, \textsc{Eagle} \citep{Schaye2015}, and \textsc{Simba} \citep{Dave2019}.
Full simulations data containing particles and groups information are publicly available and downloadable for the first four, while \textsc{Simba} data were privately provided by the authors. 

The \textsc{Illustris} and \textsc{Illustris}-TNG simulations exploit the moving-mesh code \textsc{Arepo} \citep{Springel2010}; simulated volumes and baryonic mass resolutions are as follows: 106.5$^3$ Mpc$^3$, $1.26\times10^6 M_{\odot}$ for \textsc{Illustris}; 110.7$^3$ Mpc$^3$, $1.4\times10^6 M_{\odot}$ for \textsc{Illustris}-TNG100; 302.6$^3$ Mpc$^3$, $1.1\times10^7 M_{\odot}$ for \textsc{Illustris}-TNG300. The \textsc{Eagle} simulations exploit \textsc{Gadget-3}, the the latest incarnation of the original Tree-SPH (Smoothed Particle Hydrodynamics) code developed by \citet{Springel2005}; we used data from the RefL0100N1504 simulation, which is the most complete in the suite of runs in terms of physics included in the code; the run simulates a volume of 100.0$^3$ Mpc$^3$, and has a baryonic mass resolution of $1.81\times10^6 M_{\odot}$. Finally, \textsc{Simba} is based on the \textsc{Gizmo} code by \citet{Hopkins2015}, a mesh-free finite-mass hydrodynamic code which handles shocks via  Riemann solvers, with no need for artificial viscosity; the simulation also includes detailed and novel recipes for AGN feedback, and has a volume of 100$^3$ Mpc$^3$, with a baryonic mass initial resolution of $1.82\times10^7 M_{\odot}$, i.e. comparable to the TNG-300 simulation.
All the models include baryonic sub-grid physics to simulate star and black hole formation, stellar and AGN feedback, and metal enrichment; while \textsc{Illustris}, \textsc{Illustris}-TNG and \textsc{Simba} model subgrid physics from first principles, \textsc{Eagle} use empirical relations to match observed properties of galaxies at $z=0$. 
All simulations assume a Chabrier IMF, and we applied again the corrective factor of 1.75 to compare with the Salpeter IMF adopted in our SED-fitting procedure. We also paid attention to consider the cosmological factor $h=H_0/100$ in the computation of masses and SFRs, when necessary (some simulations define masses and lengths in units of $h$, others do not).

For the models, we considered simulated galaxies with $M_*/M_{\odot}>5\times10^9$, and we adopted the usual selection criterion sSFR$<10^{-11}$ yr${^-1}$. 
Since in the \textsc{Illustris} and \textsc{Illustris}-TNG data releases multiple values of masses and SFRs are given for each simulated object, depending on the different radii within which they are computed, we tried to mimic as accurately as possible the observational approach: to this aim, we used the values estimated within twice the half mass radius of each object \citep[see also the analysis in][]{Donnari2019}. On the other hand, \textsc{Eagle} outputs masses and SFRs within a set of apertures (defined as the diameters of spheres centered on the position of the object) of fixed proper lengths, from 1 to 100 proper kpc; and two estimations of the half mass radius, $R_{1/2,30}$ and $R_{1/2,100}$, computed within 30 and 100 kpc respectively. To obtain a fair comparison with the \textsc{Illustris} and TNG cases, in this case we compute the sSFR by considering, for each simulated galaxy in a snapshot, the mass and SFR within the aperture which is closest to $4 \times R_{1/2,100}$. Finally, the values of the sSFR within twice the half mass radius of each object for \textsc{Simba} were directly communicated by the authors.

In Figure \ref{numdens} we summarize the results of our analylsis on number densities. We show three estimates from other studies on observed data (empty squares): \citet{Muzzin2013} use the color-color $UVJ$ diagram to select quiescent galaxies in COSMOS/UltraVISTA, in 7 redshift bins up to $z\sim4$ (we show the values corresponding to the $M_*/M_{\odot}>10^{10}$ selection, in grey); \citet{Straatman2014} select quiescent sources in the ZFOURGE survey, again using the $UVJ$ criterion, and focusing on the mass range log$(M/M_{\odot}) > 10.6$ and the redshift range $3.47 < z < 4.25$ (in red - but we report their lower redshifts estimates as well); and \citet{Schreiber2018b} estimate the number density of quiescent galaxies at $3<z<4$ and $M_*/M_{\odot}>3\times10^{10}$ by combining the $UVJ$ selection on ZFOURGE photometric data with MOSFIRE spectral analysis, and we plot their two estimates for strictly color-selected objects (large empty cyan square) and sSFR-threshold candidates, which include recently quenched galaxies, as in M18 (smaller cyan square). 
We then plot the results based on our SED-fitting method for the ``reference'' (both before and after the correction for incompleteness, as circles or diamonds, respectively) and the ``lines'' selections (as stars), considering both the cuts in mass and redshift adopted by S14 (red empty symbols), and the $3<z<5$ interval without any mass cut (dividing in two redshift bins, $3<z<4$ and $4<z<5$ - black solid symbols). 
Finally, we plot the results from the five hydrodynamical simulations (solid colored lines), as described above, showing also a case in which we consider the total sSFR of the simulated galaxies to select the passive ones, rather than the sSFR of the central regions, for the sake of comparison. 

Looking at the resulting plot, at $3<z\leq4$ we find a reasonable agreement between our number density estimations, the ones obtained by other observational studies, and the \textsc{Eagle} and TNG-100 models; TNG-300 is close enough as well, although its number density is slightly lower, perhaps because of the lower resolution of the simulation with respect to TNG100. Remarkably, while the original \textsc{Illustris} simulation is not able to reproduce the properties of the galactic populations at high redshift and falls short at all epochs above $z>1$ in reproducing the number of quenched galaxies, the TNG runs have largely cured this issue. On a side note, we point out that if we only considered the CANDELS fields yielding the highest densities the agreement both with other observations and with the simulations would be much less satisfying.
We also see that, consistently with what we found in M18, our estimate is lower than the one by S14 using their cuts for mass and redshift, and we ascribe this to their shallower selection criteria, which include also mildly star-forming objects in the selection.

On the other hand, at $z>4$ the TNG models still show a clear tension with the observational data, whereas the \textsc{Eagle} model performs better; however, the \textsc{Eagle} trend is too flat and struggles at reproducing the observed number densities at $z<3$. It is difficult to trace back the origin of this different trends to the various physical mechanisms in the models, because hydrodynamical simulations are highly non-linear by construction. It is worth pointing out that the \textsc{Illustris} and \textsc{TNG} mass functions are typically close to the observed ones \citep[e.g.][]{Genel2014,Pillepich2018}, pointing to the conclusion that the discrepancy at $z>4$ is not due to a poor sampling of the whole high-redshift galaxy populations, but more likely to enduring issues with the modeling of quenching mechanism. Gas hydrodymancs, radiative cooling, star formation and feedback from stars and AGNs are indeed implemented in different ways in TNG and \textsc{Eagle}. We speculate that details of the AGN feedback implementation play a major role in regulating the activity of the simulated galaxies, and therefore in the defining the properties of the passive populations at different redshifts. In particular, the thermal feedback implemented in \textsc{Eagle} is more efficient at high redhisft, while in TNG the most effective mechanism is the kinetic feedback, which has a larger impact at low redshifts \citep{Weinberger2018}, consistently with the global trends we have found. However, these are basic speculations. A detailed work specifically dedicated to the analysis of this topic is currently in the making (Fortuni et al., in preparation).


We point out that the values of SFR and masses that we choose to consider play a major role in reconciling models and observations: indeed, for TNG and \textsc{Eagle} the sSFRs computed over limited central areas turn out to be typically lower than those computed considering the full extension of the objects, and in some cases they are exactly zero, thus allowing the inclusion of more sources in the passive selection. 
On the other hand, using the simplest approach of summing on all the particles of a simulated galaxy would lead to strong tensions at all redshifts, as in manu cases the sSFRs would be higher, excluding objects from the selection criteria and finally resulting in lower number density estimates (see e.g. the blue dashed line in Fig. \ref{numdens}, which shows the case for TNG100 model). Note that this implies that in the models the stellar mass profiles are more centrally concentrated than SFR profiles (at least at high-$z$), so that a significant amount of star formation takes place in the outskirts of the galaxies \citep[see also][]{Donnari2019}: therefore, enlarging the radius in which the SFR and mass are computed yields higher sSFR values that eventually exclude many objects from the passive selection. If this is the case, current observations might underestimate the actual cosmic SFR, missing some amount of peripheral activity.

Finally, the number densities in \textsc{Simba} are closer to the observed values than the ones from \textsc{Illustris}-1, but fall short with respect to TNG and \textsc{Eagle}. Remarkably, in this model the values remain almost unchanged varying the aperture radius over which SFRs and masses are computed, at variance with the other simulations. Apparently, the peripheral star formation activity is not present in \textsc{Simba}. This might be due to the different star-formation prescriptions in the codes: while \textsc{Illustris} and TNG assume that stars can form at gas densities $n>0.13$ cm$^{-3}$ \citep{Vogelsberger2013}, in \textsc{Simba} a subgrid criteria based on the estimated local density of the H2 molecule is adopted \citep{Dave2019}, and this tends to require significantly global higher densities to trigger the activity (Dav\'e, priv. comm.). As a result, TNG might form more stars in the outskirts of galaxies than \textsc{Simba}, perhaps in particular as a result of shock-induced cooling and collapsing of gas. On the contrary, the resolution of the simulations seem to play a minor role: while the TNG-100 and \textsc{Eagle} ones are finer ($\sim 10^6 M_{\odot}$ per gas particle), the TNG-300 one is comparable to \textsc{Simba} ($\sim 10^7 M_{\odot}$ per gas particle), but its number density of passive sources is close to the one in TNG-100.

As final remarks, it must be pointed out that: (i) we are applying demanding criteria which only include in the sample very robust candidates; (ii) since we are working with the CANDELS catalogs, we are not considering $H$-dropouts; $K$ or IRAC-detected sources might increase the actual number of high-redshift passive galaxies, since sources that quenched long before the observation have negligible UV rest-frame flux, thus failing the $H$-band detection at high redshifts; (iii) as shown in Sect. \ref{completeness}, the observed values are most likely far from completeness, and although we also considered ``corrected'' values, we might anyway still be underestimating the total number of real passive sources. These points suggest that 
the tension with models cannot be considered completely ruled out. 

We point out that the volumes of the simulations are comparable to the two considered cosmological volumes for $3<z<4$ and $4<z<5$, which are of the order of $3\times10^6$ Mpc$^3$; in particular, while TNG-100, \textsc{Eagle} and \textsc{Simba} are a factor of $\sim$3 smaller than the observed volumes, TNG-300 is a factor of $\sim$9 larger, and is therefore statistically significant.

\section{The contribution of the passive population to the universal Star Formation Rate Density} \label{sfrd}

The SED-fitting approach that we adopted in this study has another advantageous outcome. The best-fit template of each galaxy can be used to infer the complete description of its SFH through the cosmic history, by means of two of its free parameters: $\tau$ (the $e$-folding time for the exponentially declining models, or the burst duration for the top-hat models) and $C$ (the normalization factor of the solution). They are the only two parameters needed to constrain the functional form of the SFH, $\Psi(z)$. From it, we can compute the average SFR of a galaxy in any redshift interval, and then obtain an estimate of the universal SFRD summing up all the contributions from individual galaxies\footnote{Starting from the fitted quantities, the procedure is as follows. Each source's best-fit model is characterized by the stellar mass $M_*$ at the age the galaxy has when it is observed, $t_f$. From these two fitted quantities, one can infer the normalization factor $C$ as
\begin{equation}
C = \frac{ M_*}{\int_0^{t_{f}} f_*(t) \times \Psi'(t) dt}, \label{sfr1}
\end{equation}
\noindent where $f_*(t)=1-f_{\mbox{rec}}(t)$ is the dimensionless parameter that accounts for the recycled fraction of gas, and $\Psi'(t)$ is the instantaneous \emph{normalized} SFR (given by the SFH model, for example $\Psi'(t)=\textrm{exp}(-t/\tau)$ for the exponentially declining models). The true instantaneous SFR at any epoch $t$ is then $\Psi(t) = C \times \Psi'(t)$.}.

\begin{figure*}
\centering
\includegraphics[width=16cm]{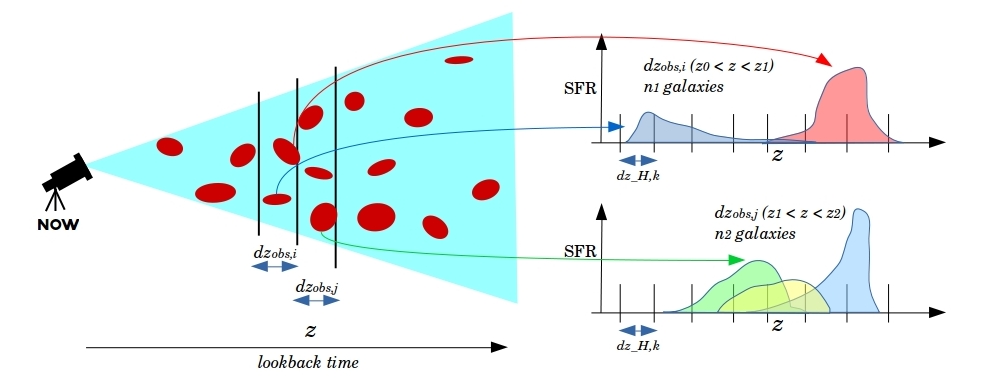}
\caption{Sketch of the method adopted to compute the global SFRD across the cosmic time. We sum the contributions of all the galaxies observed in a given redshift interval $dz_{\textrm{obs}}$ to the SFR in each redshift bin $dz_{H}$ across the cosmic history. Such contributions can be evaluated from the shape and normalisation of the SFH of each galaxy, which in turn come from the SED-fitting procedure. Then, we divide the obtained total SFR in each bin $dz_{H}$ by the $dz_{\textrm{obs}}$ comoving volume to obtain the SFRD in that $dz_{\textrm{obs}}$. Finally, we average over the $dz_{\textrm{obs}}$ bins to obtain the SFRD of each $dz_{H}$.} \label{sfrd_method}
\end{figure*}

As a cautional remark, it is worth mentioning that this approach implies the assumption that the best template in the SED-fitting process (the one with the lowest $\chi^2$) is the ``correct'' one, while in reality many more or less plausible solutions can be associated to each source. We do not take into account the uncertainties associated to this complication. In addition to this, we should also recall that it is assumed that galaxies evolve in isolation; in other words, mergers are not considered in this approach. This is by construction inherent to the SED-fitting method. However, we do not consider this point to be invalidating: the past evolution of an object and its SFH can be caused by a number of causes/events, but the amount of stars formed per year is the same, independently from the merger history.

To assess the reliability of the method, we first compare the reconstruction of the SFRD based on the SED-fitted SFHs of the CANDELS catalogs with the observed ones; we consider the well known one by \citet[][ M14]{Madau2014}, and the one by \citet[][ Y16, obtained from a MCMC technique applied to the observed mass functions rather than from the direct summation of UV and IR fluxes at the observed redshifts]{Yu2016}. For the CANDELS data, we use the delayed $\tau$-models (in which the star formation is parametrized as $SFR(t)=t^2 / \tau \times \mbox{exp}(-t/\tau)$) and the mass estimates with method 6a\_del$\tau$, presented in \citet{Santini2015}.

We point out that we cannot apply any correction for incompleteness here, since the approach is based on the computation of individual SFHs, which by definition cannot be inferred for undetected objects. In this exercise, the analysis is necessarily limited to galaxies with $H160<27.5$, i.e. to a certain mass cut, while the SFRD is typically computed by integrating the observed luminosity and mass function down to very faint limits in SFR and mass, far below the observed one, by deep extrapolation. Therefore, our estimation is actually a lower limit.

We proceed as follows. First, we define two grids of redshift bins:
\begin{itemize} 
\item $dz_{\textrm{obs}}$, which we use to bin the observed catalogue. Each bin can represent a slice of Universe which evolves comovingly, and therefore its global SFH can be considered an independent realization of an universal SFH, for $z>z_{\textrm{obs}}$;
\item $dz_{H}$, which we use to bin the cosmic history: we will obtain the global SFH of each $dz_{\textrm{obs}}$, as the sum of the SFRs of individual galaxies (computed as described above) in each $dz_{H}$.
\end{itemize}
In other words, we select from the CANDELS catalogs the galaxies observed in a bin $dz_{\textrm{obs}}$, and we trace back their SFH, summing all their contributions in each $dz_{H}$ bins. 
To obtain the SFR \emph{density}, we then divide the result by the comoving volume of $dz_{\textrm{obs}}$: i.e., the volume of the spherical shell centered on the observer, extending from $z_{\textrm{obs}}$ to $z_{\textrm{obs}}+dz_{\textrm{obs}}$, and re-scaled to the total angular area of the five fields. We then repeat the procedure for all the $dz_{\textrm{obs}}$; finally, in each $dz_H$ we compute the mean of the various SFRDs corresponding to each $dz_{\textrm{obs}}$. In formulae:

\begin{equation}
    \forall dz_{\textrm{obs}} \longrightarrow SFR_{dz_{H}} = \sum_{i=1}^n SFR_{dz_{H},i} 
\end{equation}

\noindent where $n$ is the number of galaxies in the considered $dz_{\textrm{obs}}$; then,

\begin{equation}
    \forall dz_{\textrm{obs}} \longrightarrow SFRD_{dz_{H}} = SFR_{dz_{H}} / V_{dz_{\textrm{obs}}}
\end{equation}

\noindent and finally

\begin{equation}
    \langle SFRD_{dz_{H}} \rangle = \frac{\sum_{j=1}^{N_{dz_{\textrm{obs}}}} SFRD_{dz_{H},j}}{N_{dz_{\textrm{obs}}}}
\end{equation}

\begin{figure}
\centering
\includegraphics[width=8.5cm]{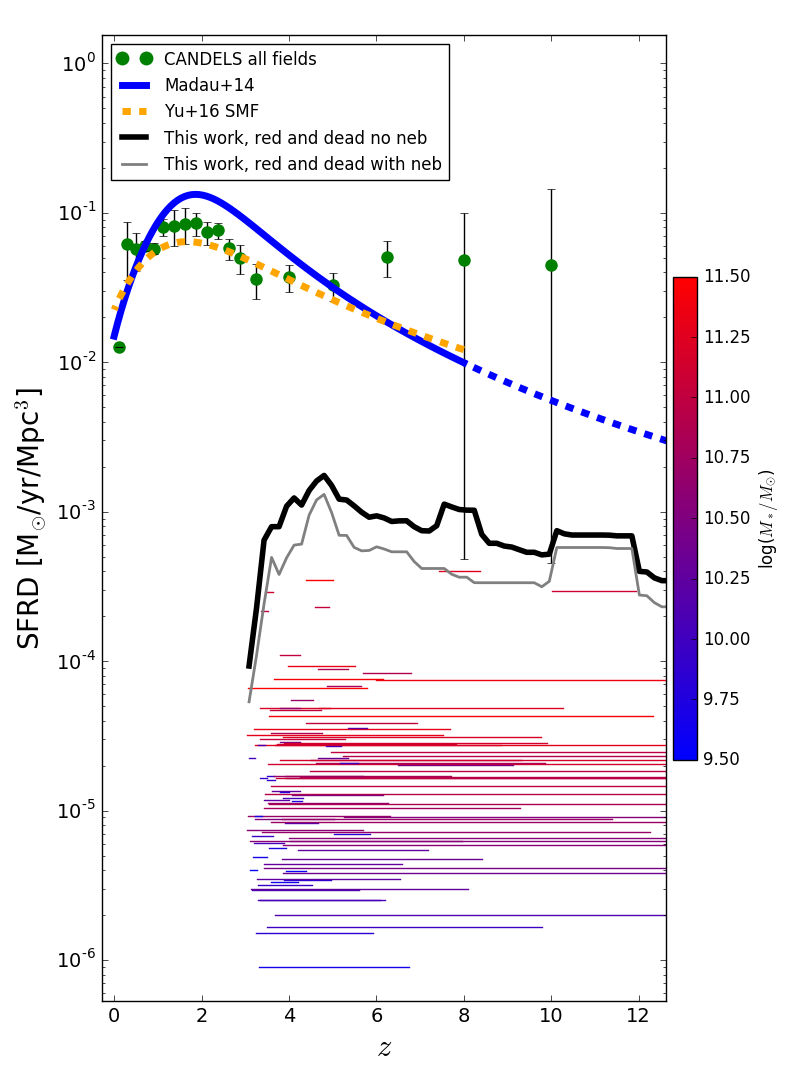}
\caption{Universal star formation rate densities. The blue solid thick line is the \citet{Madau2014} SFRD inferred from UV and FIR observational data. The dashed yellow thick line is the SFRD by \citet{Yu2016}, who infer the instantaneous SFR from the stellar mass functions at various redshifts by means of a MCMC approach, and argue that a substantial offset from the observed curve is evident. The green dots are our determination of the SFRD using the five CANDELS fields 
as described in the text: the mean of the SFRDs at all epochs obtained from samples of objects at various observed bins of redshifts $dz_{\textrm{obs}}$ (the error bars represent the standard deviations of the distribution of the SFRD from each $dz_{\textrm{obs}}$, see text for definitions). The points are in good agreement with observations (particularly with the Yu et al. curve), reassuring that the method is reliable. 
Finally, the black and grey solid lines are the SFRDs of our selections of passive galaxies in the CANDELS fields (respectively with and without nebular emission lines), and the thin blue to red horizontal lines are the individual contribution of the candidates to the SFRD, color-coded as a function of their observed mass.}
\label{SFRDtotal}
\end{figure}

The method is sketched in Figure \ref{sfrd_method}. In Fig. \ref{SFRDtotal} we plot the result of this approach, together with the two cited observational ones.

Our points are in good agreement with the curves, and particularly with Y16, although they are consistent with both estimations considering the uncertainties (the error bars are the standard deviations of the distribution of the SFRD from each $dz_{\textrm{obs}}$). We speculate that while the discrepancy with the M14 curve can be understood in terms of the caveats on the incompleteness of our sample discussed above, the good consistency at $z<5$ with the Y16 estimate is justified by the fact that we are basically adopting their same method to compute the SFRD (see their Eq. 5). 
Therefore, the SFR estimated from the observed UV/IR fluxes seems not strictly equivalent to the SFR inferred from the observed stellar mass, which might be due to several factors discussed in Y16, among which the IMF, the metallicity, the outshining of young stars, or overestimated absorption, as well as possible differences in the computation of the recycled fraction. However, the tension with MD14 is not dramatic, 
and furthermore, we note that the precise values of the averaged SFRDs depend on the choice of  $dz_{\textrm{obs}}$ binning (and \textit{binning is sinning}...).

At $z>5$ our estimation apparently diverges from the other two, but the scatter between the SFRDs from the various $dz_{\textrm{obs}}$ bins is large enough to make the global estimate consistent with the curve within the error budget. We recall that the bars correspond the standard deviation of the distribution of the individual SFRDs of each $dz_{\textrm{obs}}$, so that their extension at high $z$ is likely due to poor statistics (few detected objects, plus cosmic variance); on the other hand, at low redshifts the $dz_{\textrm{obs}}$ bins contributing to the summation are less, so their dispersion is lower. 

Having assessed that this approach can reasonably reproduce known observational constraints, we can now exploit it to estimate the contribution to the cosmic SFRD of the galaxies that we have selected as passively evolving at the time of their observation, during their previous phase of SF activity. Applying the very same technique to our sample (taking $dz_{\textrm{obs}}=3-5$, and considering the top-hat SFHs we used in the SED-fitting procedure rather than the standard $\tau$-models) we obtain the thick black (for the ``reference'' selection, without nebular emission lines) and thin grey (for the ``lines'' selection) curves plotted in Fig. \ref{SFRDtotal}. We note that the sharp decrease in the plotted SFRD of passive galaxies towards $z=3$ is of course an artifact due to the fact that we are only looking for $z>3$ objects, so all our candidates cease to form stars at earlier epochs, by construction. 
In this case, we must recall that as discussed in Sect. \ref{completeness} we are probably missing passive sources due to our demanding passive selection criteria, on top of the mentioned photometric incompleteness. As already discussed, we cannot apply any correction here, so again we must take the result as a lower limit of the true values.

Considering the ``reference'' sample, it turns out that the contribution of the passive candidates to the global SFRD is $5-10\%$ of the total, at $z<8$. We can compare this result to their fractional abundance in number: from our estimates summarized in Tab. \ref{selects}, they are $102/21608\simeq 0.5\%$ of the total number of all $z>3$ detected sources. Their SF activity must therefore be $\sim10-20$ times higher than the average.

In Fig. \ref{SFRDtotal} we also plot the contribution to the SFRD from the individual SFHs of the passive candidates (thin lines, color coded from blue to red according to their final stellar mass; they are straight horizontal lines, because the SFR is constant for the top-hat models). A few objects have a very high impact on the total SFRD and this is reflected in the sudden jumps in the the global curve. However, the bulk of the contribution is given by longer, intermediate intensity bursts. 

As a final note we also remark that the most massive objects (redder lines) typically have higher SFRs. Since the lines are plotted as a function of redshift, it might not be immediate to realize that the actual duration of the bursts of SF activity are not dramatically different (e.g., from $\simeq1.9$ Gyr for a burst starting at $z=20$ to $\simeq1.2$ Gyr for one starting at $z=6$, both ending at $z=3$). This clarifies the direct correlation between the mass of the galaxies (traced by the color of the lines) and the value of their SFRD. Nevertheless, it is worth noticing that the galaxies with the highest SFRDs ($>10^{-4}$ $M_{\odot}$/yr/Mpc$^3$) have very short bursts, implying an extremely fast and efficient star formation activity abruptly quenched either by gas consumption or by a very effective feedback mechanism.




\begin{figure*}
\centering
\includegraphics[width=14cm]{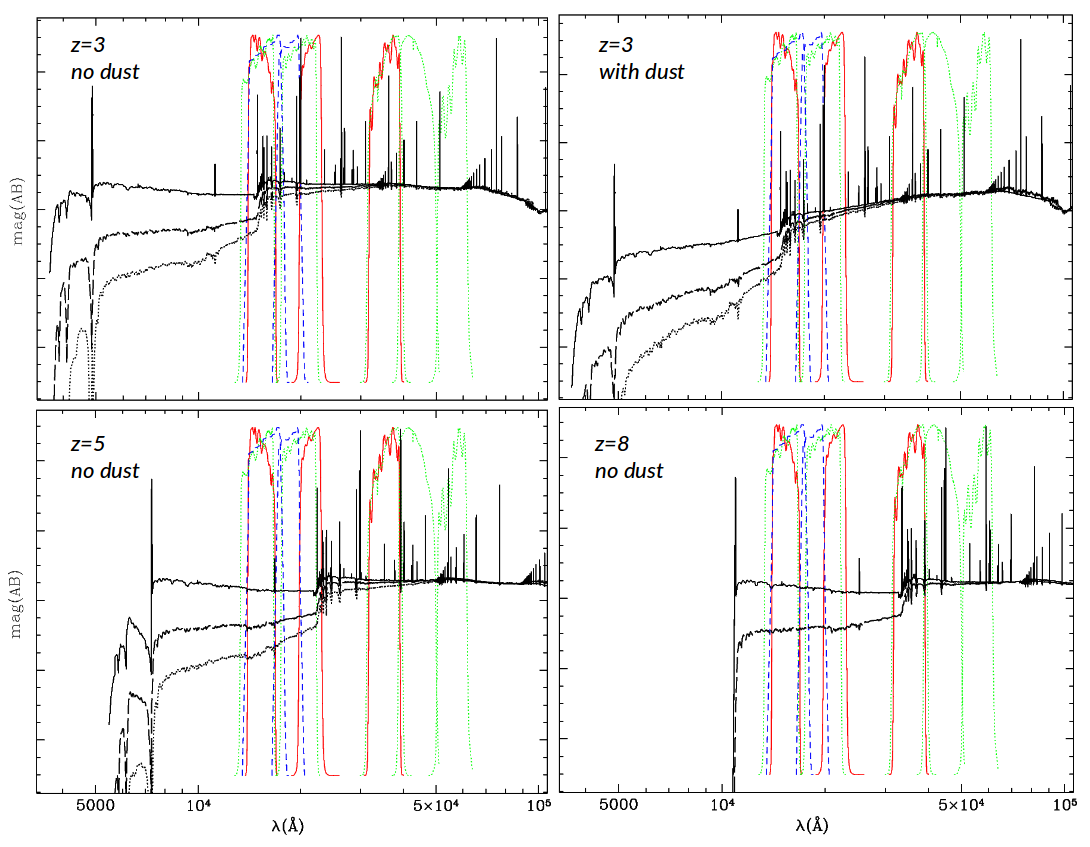}
\caption{Compared capabilities of present and future facilities in detecting passive sources at different redshifts. The black lines are the SEDs of galaxies that quench at different epochs (redshifts), as indicated in the panels (the magnitude scales on the $y$ axis are arbitrary). The SEDs were simulated with the top-hat library used in this paper. The solid line in each panel, which has brighter UV fluxes, shows the SED at the epoch of quenching $t_{\textrm{quench}}$, while the dashed and dotted lines with fainter and fainter UV fluxes correspond to 50 Myr and 300 Myr later, respectively. The $z=8$ panel does not include the $t_{\textrm{quench}}+300$Myr case, because it is not modeled in our libraries. 
Note that the upper right panel shows the SEDs of a galaxy with dust ($E(B-V)=0.2$), while the other cases are dust-free: the UV region of its spectrum at $t_{\textrm{quench}}$ epoch is very similar to the $t_{\textrm{quench}}+50$Myr epoch of the corresponding dust-free case, a known degeneracy that can be broken sampling the $D_{4000}$ break, obtaining the Mg$_{UV}$ spectral index, or looking at the FIR region of the spectra.
Superposed to the SEDs in arbitrary units are the the curves of some significant pass-band filters of CANDELS (solid red: left to right, F160, Ks and IRAC-CH1), WFIRST (dashed blue: left to right, H158 and F184), and JWST (dotted green: left to right, NIRCAM F150, F200, F356, F444, and MIRI F560). See text for discussion.}
\label{modelsseds}
\end{figure*}

\section{Summary and conclusions} \label{conclusions}

Constraining the number and the features of high-redshift passive galaxies is crucial to understand the evolution of the Universe. To this aim, we have searched for red and dead objects at $z>3$ in the five CANDELS fields, improving upon the first work on GOODS-South presented in M18. We performed SED-fitting with ad-hoc, abruptly quenched (top-hat) SFHs rather than standard $\tau$-models, and exploited a probabilistic technique to select robust candidates considering all the possible solutions of the fit and only keeping sources with zero or very low probability star-forming ones ($p_{\textrm{SF}}<5\%$). 

Our selection method allows us to find galaxies that would not be identified as passive with more traditional techniques such as the widely used $UVJ$ color-color diagram. In particular, we can single out objects that fail the $U-V$ color selection, having abruptly quenched their SF activity only shortly before the observation. The quenching might be caused by a strong quasar activity episode, which typically happens on very short timescales; moreover, feedback from young stars in the form of UV radiation, winds and supernovae can prevent gas from infalling back to the galaxy avoiding re-juvenation processes, at the same time preventing the local gas (which has been observed to exist in quenched galaxies) to cool down and collapse into new stars.
Other possible explanations might include low nuclear activity, shock heating of infalling gas, strangulation/removing of gas, or low SF efficiency. The relative compactness of the observed passive sample (Fig. \ref{mr}) hints at an inside-out formation process, in which peripheral star formation due to minor mergers might happen at lower $z$.

Depending on the chosen library of models and SED-fitting technique, we end up with the selections summarized in Table \ref{selects}. We individuate 102 red and dead candidates, which decrease to 40 if we  include the emission lines in the libraries. 
The physical properties of all the candidates, as inferred from their best fit, are given in Table \ref{tab1}; snapshots and SEDs are available on-line as complementary material. 

We find the number densities given in Tabs. \ref{tnd} and \ref{tnd2}, with a global estimation of $\simeq1.73 \pm 0.17 \times10^{-5}$  ($6.69 \pm 1.08 \times 10^{-6}$) Mpc$^{-3}$ at $3<z\leq5$, for the ``reference'' (``lines'') selection. The estimate rises to $\simeq2.30 \pm 0.20 \times10^{-5}$ after multiplying by a correction factor to take into account the estimated incompleteness of our selected sample.
We compare these numbers with the ones obtained by five cosmological hydrodynamical simulations, for which we adopt the $sSFR<10^{-11}$ yr$^{-1}$ threshold to identify passive galaxies with $M_*/M_{\odot}>5\times10^9$. We find a reasonable agreement with the predictions from the \textsc{Illustris}-TNG and \textsc{Eagle} models at $3<z\leq4$, provided that a wise choice of the relevant data is made, i.e. including in the computations only the stellar particles within some typical radius reminiscent of an observed quantity (e.g. twice the half mass radii), rather than summing on all the particles formally belonging to a simulated galaxy (if we do so, many objects formerly passing the selection would instead be classified as star forming). Interestingly, this would indicate that the observations might be missing to detect some level peripheral star formation activity.
However, a non negligible tension remains at higher redshifts, where all the models fail to produce a significant number of quenched objects, whereas we find 14 galaxies in the redshift bin $4<z\leq5$. The other models (\textsc{Simba} and \textsc{Illustris}-1) provide less accurate matching at all epochs. An important remark is that some underestimation of the real number density of passive sources is probably present and unavoidable, due to many factors (e.g., the demanding selection criteria, the conservative corrections for incompleteness, the atypical number density of the COSMOS field which lowers the average), so that we cannot claim that the tension between models and observation is ruled out.


Finally, we computed the contribution of passive galaxies, as inferred from our selection, to the global cosmic SFRD during their past active phase, finding that despite being a tiny fraction ($\sim0.5\%$) of the whole galactic population, they provide a $\sim 5-10\%$ fraction of the total activity from $z\simeq8$ to $z\simeq3$. Therefore, they play a crucial role in the cosmic star formation history.

As discussed in M18, upcoming observations with telescopes of the new generation will provide more stringent constraints on the observational data. In Fig. \ref{modelsseds} we give an example of the current and future possibilities. We plot, with arbitrary units, the SEDs of top-hat models at $z=3$, 5 and 8, each one at three different epochs: the time of the quenching of star-formation $t_{\textrm{quench}}$, 50 Myr later, and 300 Myr later (the latter except for the $z=8$ case since it is not included in our library of models). The models are dust-free, but for comparison we also show the case of the $z=3$ models with $E(B-V)=0.2$: its rest-frame spectrum blueward of the $4000 \textrm{\AA}$ ($D_{4000}$) break at $t_{\textrm{quench}}$ is very similar to the spectrum at $t_{\textrm{quench}}$+50Myr of the corresponding dust-free model. This creates a degeneracy that can be broken with a better sampling of the break, looking at FIR wavelengths, or using well tailored spectral indexes such as Mg$_{UV}$ \citep{Daddi2005}.

As it can be seen looking at the leftmost red filter curve, $HST$ $H$160, the current work is based on a NIR selection which corresponds to UV/optical rest-frame at $z>3$. Since soon after the quenching of the star formation activity the luminosity drops by 1-2 magnitudes in this range of wavelenghts, current selections are most likely biased towards recently quenched sources at $z>4$, where the detection falls blueward of the break: at such redshifts we are probably missing galaxies that have quenched more than a few tens of Myr before the observation epoch.
On the other hand, $JWST$ will straddle the optical/NIR rest-frame wavelengths with the highest resolution, sampling the $D_{4000}$ break for $z\sim8$ sources with the F356 NIRCAM filter (which corresponds to the IRAC-CH1 bandpass), at the same time enhancing the SNR reaching AB magnitudes $\sim29-30$ and the resolution of the images. Furthermore, while we are now forced to work on a very restricted sky area, WFIRST will observe a $\sim2500$ sq. degrees reaching $j\sim27$ (29 in the deep survey), allowing for much grander statistic significance, although its filter coverage is similar to the CANDELS $HST$ passbands so that it will likely be useful only for detection.

In sinergy with FIR facilities like ALMA, which can confirm the nature of the candidates analyzing the gas and dust content as we did in S19, or EELT which will provide spectral data for extremely faint objects (together with robust morphological analysis thanks to adaptive optics technology), the accuracy and statistical significance of data will quite soon be enormously enhanced. This will hopefully allow to finally understand whether the currently determined numbers, which the models seem to be able to broadly reproduce up to $z\simeq4$ but fail to match at earlier epochs, are indeed a good estimation of the real number of passive objects, or they are actually lower limits - in which case a tension with the theoretical predictions would reappear even at lower redshifts.


\section*{Acknowledgements}
We acknowledge the Virgo Consortium for making their simulation data available. The \textsc{Eagle} simulations were performed using the DiRAC-2 facility at Durham, managed by the ICC, and the PRACE facility
Curie based in France at TGCC, CEA, Bruyèresle-Châtel; we would like to thank Claudio Dalla Vecchia for the kind help with the data handling. We also thank Romeel Dav\'e for providing the \textsc{Simba} simulation data and for his help in interpreting them. 




\bibliographystyle{mnras}
\bibliography{biblio} 




\appendix
\onecolumn
\section{The TOP-HAT library} \label{THlib}

The library used to fit the observed photometry with \textsc{zphot} consists of $\sim 7.9 \times 10^8$ models. 
The grid of models is built considering the following criteria:
\begin{itemize}
    \item we assume a \citet{Salpeter1959} IMF;
    \item redshifts vary from 0.0 to 8.0. These are the redshift at which the objects are observed, and at which the templates are shifted applying the K-correction;
    \item the burst durations $\Delta t_{\textrm{burst}}$ vary from 100 Myr to 1.5 Gyr (with linear steps of 100 Myr); note that at $z=3$ the Universe has an age of $\simeq 2.2$ Gyr;
    \item we considered two extinction laws, namely the \citet{Calzetti2000} and the Small Magellanic Cloud by \citet{Prevot1984} ones;
    \item the ages and extinctions values vary as a function of $\Delta t_{\textrm{burst}}$. We considered three possible regimes:
    \begin{itemize}
        \item for $t \leq \Delta t_{\textrm{burst}}$, we built a model for each 50 Myr of age, and allowed for values of $E(B-V)$ from 0 to 1;
        \item for the first 300 Myr after the quenching, we again built a models for each 50 Myr of age, but restricted the allowed values of $E(B-V)$ to the range $0-0.4$;
        \item for later ages, we built models for each 50 Myr of age up to 1 Gyr after the quenching, and then we used logarithmic steps of 0.1 to reach the age of the Universe (at $z=2$, again to allow for more relaxed fitting), and considered $E(B-V)<0.2$.
    \end{itemize}
    These choices for the $E(B-V)$ values were made to mimic the expected drop of dust content in a quenched galaxy after the end of the star-forming activity.
    \item the emission lines are included using the prescription described in \citet{Castellano2014,Schaerer2009}.
\end{itemize}

\newpage 
\section{Physical properties of the selected sample of red and dead candidates} \label{phystab}
\renewcommand{\arraystretch}{1.5}
\begin{longtable}{ | l || l | l | l | l | l |}
\hline
ID$_{\textrm{CAND}}$ & RA \space DEC [deg] & $z_{\textrm{CAND}}$ & Age [Gyr] & $M_*$ [$10^9$ M$_{\odot}$] &  $\Delta t_{\textrm{burst}}$ [Gyr] \\ \hline\hline

COSMOS-16676* $\oplus$ & 150.0615 \space\space 2.3787 & 3.72 & $0.50_{-0.15}^{+0.65}$ & $322.70_{-165.00}^{+153.3}$ & 0.2    \\ \hline
COSMOS-19502 $\oplus$ & 150.1309 \space\space 2.4136 & 3.87 & $0.40_{-0.15}^{+1.00}$ & $107.40_{-31.04}^{+73.20}$ & 0.1    \\ \hline
COSMOS-2075 & 150.0535 \space\space 2.2045 & 3.35 & $1.00_{-0.99}^{+0.70}$ & $45.59_{-32.39}^{+29.71}$ & 0.9    \\ \hline
COSMOS-18286 & 150.0767 \space\space 2.3986 & 3.04 & $1.40_{-1.15}^{+0.25}$ & $199.40_{-89.30}^{+72.30}$ & 1.4    \\ \hline\hline

EGS-14727 $\oplus$ & 214.8956 \space\space 52.8566 & 3.05 & $0.90_{-0.60}^{+1.00}$ & $155.90_{-60.74}^{+67.30}$ & 0.7    \\ \hline
EGS-21351 $\oplus$ & 214.6736 \space\space 52.7326 & 3.61 & $0.65_{-0.35}^{+0.90}$ & $80.76_{-23.84}^{+37.04}$ & 0.2    \\ \hline
EGS-24177 $\oplus$ & 214.8661 \space\space 52.8843 & 3.42 & $0.80_{-0.45}^{+0.85}$ & $214.40_{-59.60}^{+83.00}$ & 0.5    \\ \hline
EGS-25724 $\oplus$ & 214.9978 \space\space 52.9862 & 3.79 & $1.10_{-0.90}^{+0.35}$ & $77.77_{-35.44}^{+57.93}$ & 0.8    \\ \hline
EGS-29547 $\oplus$ & 214.6953 \space\space 52.7969 & 3.15 & $1.70_{-1.40}^{+0.15}$ & $86.37_{-42.06}^{+25.43}$ & 1.5    \\ \hline
EGS-2490 & 214.9515 \space\space 52.8292 & 3.10 & $1.40_{-1.25}^{+0.20}$ & $39.16_{-19.80}^{+10.60}$ & 1.4    \\ \hline
EGS-6539 & 214.9132 \space\space 52.8247 & 3.44 & $1.55_{-1.40}^{+0.10}$ & $136.50_{-52.55}^{+51.40}$ & 1.5    \\ \hline
EGS-15868 & 214.8712 \space\space 52.8451 & 3.61 & $0.30_{-0.15}^{+1.25}$ & $105.10_{-31.18}^{+80.60}$ & 0.2    \\ \hline
EGS-23036 & 214.8791 \space\space 52.8881 & 3.57 & $1.55_{-1.40}^{+0.05}$ & $36.73_{-19.49}^{+17.00}$ & 1.4    \\ \hline
EGS-24356 & 214.6201 \space\space 52.7096 & 3.43 & $1.65_{-1.45}^{+0.00}$ & $109.60_{-49.35}^{+38.10}$ & 1.5    \\ \hline
EGS-26762 & 214.6261 \space\space 52.7268 & 3.28 & $1.10_{-0.90}^{+0.65}$ & $19.41_{-9.34}^{+10.78}$ & 1.  0  \\ \hline
EGS-27491 & 214.6177 \space\space 52.7242 & 3.34 & $1.65_{-1.45}^{+0.05}$ & $96.93_{-55.25}^{+51.07}$ & 1.5    \\ \hline
EGS-30675 & 214.9049 \space\space 52.9354 & 3.01 & $1.25_{-1.10}^{+0.60}$ & $50.26_{-22.48}^{+17.50}$ & 1.2    \\ \hline \hline

GOODSN-13 $\oplus$ & 189.1544 \space\space 62.0947 & 3.01 & $0.15_{-0.05}^{+1.20}$ & $11.03_{-3.20}^{+11.54}$ & 0.1    \\ \hline
GOODSN-2901 $\oplus$ & 189.2199 \space\space 62.1569 & 3.70 & $0.40_{-0.25}^{+1.10}$ & $11.65_{-3.73}^{+10.30}$ & 0.3    \\ \hline
GOODSN-4004 $\oplus$ & 189.2657 \space\space 62.1684 & 3.81 & $1.40_{-1.20}^{+0.05}$ & $36.48_{-13.20}^{+27.85}$ & 1.3    \\ \hline
GOODSN-5059 $\oplus$ & 189.1623 \space\space 62.1782 & 3.69 & $1.30_{-0.95}^{+0.20}$ & $126.70_{-54.97}^{+63.40}$ & 0.1    \\ \hline
GOODSN-10672 $\oplus$ & 189.0325 \space\space 62.2164 & 6.71 & $0.50_{-0.35}^{+0.15}$ & $40.29_{-15.96}^{+42.77}$ & 0.3    \\ \hline
GOODSN-12446 $\oplus$ & 189.3730 \space\space 62.2287 & 3.05 & $0.15_{-0.10}^{+0.35}$ & $1.94_{-0.59}^{+0.99}$ & 0.1    \\ \hline
GOODSN-13403 $\oplus$ & 189.2779 \space\space 62.2350 & 3.79 & $0.45_{-0.30}^{+1.00}$ & $6.49_{-3.06}^{+4.50}$ & 0.4    \\ \hline
GOODSN-13800 $\oplus$ & 189.3323 \space\space 62.2370 & 3.33 & $0.65_{-0.40}^{+1.05}$ & $74.62_{-28.39}^{+21.80}$ & 0.5    \\ \hline
GOODSN-15054 $\oplus$ & 189.0799 \space\space 62.2448 & 3.06 & $1.40_{-1.15}^{+0.35}$ & $208.80_{-58.00}^{+77.40}$ & 1.3    \\ \hline
GOODSN-19580 $\oplus$ & 189.4824 \space\space 62.2739 & 3.10 & $0.80_{-0.50}^{+1.10}$ & $106.40_{-37.85}^{+35.70}$ & 0.5    \\ \hline
GOODSN-24501 $\oplus$ & 189.3026 \space\space 62.3665 & 4.25 & $0.35_{-0.15}^{+0.85}$ & $20.98_{-5.01}^{+18.80}$ & 0.2    \\ \hline
GOODSN-357 & 189.1611 \space\space 62.1129 & 3.09 & $0.35_{-0.20}^{+1.15}$ & $9.52_{-2.29}^{+7.22}$ & 0.3    \\ \hline
GOODSN-1570 & 189.0083 \space\space 62.1412 & 3.23 & $0.15_{-0.00}^{+0.70}$ & $13.15_{-0.45}^{+10.67}$ & 0.1    \\ \hline
GOODSN-4691 & 189.1099 \space\space 62.1752 & 3.18 & $0.90_{-0.55}^{+0.90}$ & $108.90_{-43.43}^{+56.40}$ & 0.8    \\ \hline
GOODSN-5744 & 189.1001 \space\space 62.1836 & 3.46 & $0.50_{-0.25}^{+1.15}$ & $50.56_{-21.30}^{+35.84}$ & 0.2    \\ \hline
GOODSN-6430 & 189.1844 \space\space 62.1882 & 3.21 & $0.65_{-0.50}^{+0.40}$ & $8.79_{-3.16}^{+2.34}$ & 0.6    \\ \hline
GOODSN-6620 & 189.1814 \space\space 62.1893 & 3.70 & $0.25_{-0.10}^{+1.10}$ & $27.56_{-9.58}^{+18.88}$ & 0.2    \\ \hline
GOODSN-7385 & 188.9659 \space\space 62.1945 & 3.18 & $0.30_{-0.15}^{+1.05}$ & $7.87_{-3.71}^{+3.06}$ & 0.1    \\ \hline
GOODSN-9626 & 189.1384 \space\space 62.2095 & 3.18 & $0.20_{-0.05}^{+1.15}$ & $7.83_{-2.36}^{+6.12}$ & 0.1    \\ \hline
GOODSN-10956 & 189.1512 \space\space 62.2184 & 3.09 & $0.20_{-0.05}^{+1.30}$ & $4.33_{-0.52}^{+4.48}$ & 0.1    \\ \hline
GOODSN-11579 & 189.2347 \space\space 62.2227 & 3.17 & $1.00_{-0.65}^{+0.85}$ & $53.22_{-22.52}^{+30.88}$ & 0.7    \\ \hline
GOODSN-13007 & 189.1825 \space\space 62.2320 & 3.04 & $1.10_{-0.95}^{+0.50}$ & $37.04_{-19.58}^{+9.83}$ & 1.1    \\ \hline
GOODSN-13435 & 188.9787 \space\space 62.2350 & 3.65 & $0.70_{-0.50}^{+0.85}$ & $61.89_{-26.32}^{+34.91}$ & 0.2    \\ \hline
GOODSN-14482 & 189.3675 \space\space 62.2418 & 3.49 & $0.40_{-0.20}^{+1.10}$ & $11.27_{-2.83}^{+6.16}$ & 0.2    \\ \hline
GOODSN-16817 & 189.3778 \space\space 62.2569 & 3.69 & $0.15_{-0.00}^{+1.05}$ & $6.50_{-1.47}^{+7.48}$ & 0.1    \\ \hline
GOODSN-18860 & 189.2944 \space\space 62.2694 & 4.53 & $0.20_{-0.05}^{+0.85}$ & $12.83_{-3.33}^{+17.96}$ & 0.1    \\ \hline
GOODSN-20589 & 189.1298 \space\space 62.2812 & 3.34 & $0.20_{-0.05}^{+1.00}$ & $7.62_{-0.86}^{+5.97}$ & 0.1    \\ \hline
GOODSN-21034 & 189.3571 \space\space 62.2847 & 3.33 & $1.40_{-1.25}^{+0.30}$ & $9.52_{-4.46}^{+5.35}$ & 1.3    \\ \hline
GOODSN-21961 & 189.5017 \space\space 62.2916 & 3.36 & $1.55_{-1.40}^{+0.15}$ & $27.71_{-10.44}^{+13.54}$ & 1.5    \\ \hline
GOODSN-22398 & 189.1517 \space\space 62.2950 & 3.11 & $0.25_{-0.10}^{+0.60}$ & $4.72_{-1.33}^{+2.15}$ & 0.2    \\ \hline
GOODSN-24092 & 189.3094 \space\space 62.3801 & 3.30 & $1.25_{-0.95}^{+0.50}$ & $124.30_{-44.58}^{+39.90}$ & 1.  0  \\ \hline
GOODSN-24572 & 189.3214 \space\space 62.3529 & 3.34 & $0.35_{-0.20}^{+0.60}$ & $16.75_{-5.13}^{+6.27}$ & 0.3    \\ \hline
GOODSN-25209 & 189.4217 \space\space 62.3420 & 3.27 & $1.40_{-1.15}^{+0.35}$ & $62.06_{-29.56}^{+19.13}$ & 1.3    \\ \hline
GOODSN-27251 & 189.1960 \space\space 62.3121 & 3.12 & $0.45_{-0.30}^{+0.75}$ & $11.67_{-5.61}^{+5.65}$ & 0.4    \\ \hline
GOODSN-28344 & 189.0797 \space\space 62.1513 & 4.76 & $0.70_{-0.69}^{+0.35}$ & $27.27_{-15.78}^{+26.72}$ & 0.3    \\ \hline
GOODSN-35028 & 189.4713 \space\space 62.3227 & 3.64 & $1.10_{-0.90}^{+0.45}$ & $174.50_{-106.59}^{+90.50}$ & 0.1    \\ \hline \hline

GOODSS-2608 $\oplus$ & 53.1486 \space\space -27.8896 & 3.72 & $0.30_{-0.15}^{+0.75}$ & $3.78_{-1.17}^{+2.34}$ & 0.2    \\ \hline
GOODSS-2717 $\oplus$ & 53.1893 \space\space -27.8885 & 3.02 & $0.40_{-0.05}^{+1.55}$ & $131.10_{-37.88}^{+89.50}$ & 0.1    \\ \hline
GOODSS-2782 $\oplus$ & 53.0836 \space\space -27.8875 & 3.58 & $1.25_{-1.00}^{+0.30}$ & $72.30_{-30.09}^{+19.26}$ & 0.8    \\ \hline
GOODSS-3912 $\oplus$ & 53.0622 \space\space -27.8750 & 3.90 & $1.25_{-1.05}^{+0.15}$ & $35.16_{-16.54}^{+22.62}$ & 1.2    \\ \hline
GOODSS-4587 $\oplus$ & 53.0705 \space\space -27.8686 & 3.75 & $0.25_{-0.10}^{+1.20}$ & $5.49_{-2.01}^{+4.35}$ & 0.1    \\ \hline
GOODSS-8785 $\oplus$ & 53.0818 \space\space -27.8287 & 3.85 & $1.40_{-1.20}^{+0.00}$ & $35.57_{-17.21}^{+18.63}$ & 0.9    \\ \hline
GOODSS-9209 $\oplus$ & 53.1082 \space\space -27.8251 & 4.49 & $1.10_{-0.90}^{+0.05}$ & $91.80_{-25.15}^{+5.54}$ & 1.1    \\ \hline
GOODSS-10578 $\oplus$ & 53.1653 \space\space -27.8141 & 3.06 & $1.10_{-0.85}^{+0.30}$ & $333.60_{-138.60}^{+43.20}$ & 1.1    \\ \hline
GOODSS-17749 $\oplus$ & 53.1969 \space\space -27.7605 & 3.70 & $1.40_{-1.15}^{+0.10}$ & $96.49_{-38.16}^{+36.91}$ & 0.9    \\ \hline
GOODSS-18180 $\oplus$ & 53.1812 \space\space -27.7564 & 3.65 & $1.40_{-1.15}^{+0.15}$ & $80.81_{-25.87}^{+30.69}$ & 0.8    \\ \hline
GOODSS-19883 $\oplus$ & 53.0107 \space\space -27.7416 & 3.57 & $1.25_{-1.00}^{+0.35}$ & $156.80_{-75.91}^{+39.80}$ & 1.1    \\ \hline
GOODSS-22085 $\oplus$ & 53.0739 \space\space -27.7222 & 3.47 & $1.40_{-1.05}^{+0.25}$ & $48.20_{-16.13}^{+15.77}$ & 1.2    \\ \hline
GOODSS-23626 $\oplus$ & 53.1030 \space\space -27.7123 & 4.75 & $0.45_{-0.25}^{+0.60}$ & $77.19_{-23.69}^{+38.31}$ & 0.2    \\ \hline
GOODSS-3718 & 53.1593 \space\space -27.8772 & 3.85 & $1.00_{-0.85}^{+0.40}$ & $21.51_{-9.09}^{+20.85}$ & 1.  0  \\ \hline
GOODSS-3897 & 53.0554 \space\space -27.8753 & 3.12 & $1.40_{-1.25}^{+0.10}$ & $19.12_{-9.58}^{+1.82}$ & 1.4    \\ \hline
GOODSS-3973 & 53.1393 \space\space -27.8745 & 3.63 & $1.25_{-0.90}^{+0.30}$ & $170.00_{-69.10}^{+82.70}$ & 0.8    \\ \hline
GOODSS-4202 & 53.1881 \space\space -27.8725 & 3.31 & $1.10_{-0.90}^{+0.40}$ & $4.53_{-2.43}^{+1.56}$ & 1.1    \\ \hline
GOODSS-4503 & 53.1133 \space\space -27.8699 & 3.59 & $1.25_{-1.00}^{+0.30}$ & $139.00_{-55.82}^{+57.50}$ & 1.1    \\ \hline
GOODSS-4949 & 53.0947 \space\space -27.8651 & 4.83 & $0.20_{-0.05}^{+0.85}$ & $10.08_{-3.36}^{+6.25}$ & 0.1    \\ \hline
GOODSS-5934 & 53.1295 \space\space -27.8550 & 4.86 & $0.25_{-0.10}^{+0.80}$ & $6.73_{-1.95}^{+6.25}$ & 0.2    \\ \hline
GOODSS-6407 & 53.0778 \space\space -27.8501 & 4.81 & $0.25_{-0.10}^{+0.80}$ & $16.85_{-4.34}^{+12.93}$ & 0.1    \\ \hline
GOODSS-7526 & 53.0787 \space\space -27.8395 & 3.32 & $1.00_{-0.80}^{+0.75}$ & $34.40_{-18.18}^{+15.19}$ & 0.6    \\ \hline
GOODSS-7688 & 53.0796 \space\space -27.8382 & 3.40 & $1.65_{-1.50}^{+0.00}$ & $23.70_{-12.16}^{+10.64}$ & 1.4    \\ \hline
GOODSS-8242 & 53.0816 \space\space -27.8334 & 3.24 & $1.00_{-0.50}^{+0.50}$ & $6.91_{-1.79}^{+1.26}$ & 1.  0  \\ \hline
GOODSS-12178 & 53.0393 \space\space -27.7993 & 3.29 & $1.55_{-1.35}^{+0.20}$ & $47.44_{-19.33}^{+16.17}$ & 1.5    \\ \hline
GOODSS-13394 & 53.0705 \space\space -27.7909 & 3.29 & $1.00_{-0.60}^{+0.50}$ & $11.36_{-2.83}^{+6.49}$ & 1.  0  \\ \hline
GOODSS-15457 & 53.1483 \space\space -27.7784 & 3.50 & $0.35_{-0.20}^{+0.70}$ & $4.69_{-1.53}^{+3.57}$ & 0.3    \\ \hline
GOODSS-16506 & 53.1618 \space\space -27.7706 & 3.38 & $0.30_{-0.15}^{+0.75}$ & $5.23_{-1.38}^{+3.82}$ & 0.2    \\ \hline
GOODSS-16526 & 53.0276 \space\space -27.7703 & 3.15 & $1.00_{-0.60}^{+0.05}$ & $13.57_{-4.08}^{+1.24}$ & 1.  0  \\ \hline
GOODSS-19301 & 53.1320 \space\space -27.7468 & 3.59 & $1.55_{-1.40}^{+0.00}$ & $13.42_{-7.21}^{+7.83}$ & 1.5    \\ \hline
GOODSS-19446 & 53.1652 \space\space -27.7458 & 3.27 & $1.10_{-0.90}^{+0.40}$ & $17.60_{-7.02}^{+4.90}$ & 1.1    \\ \hline
GOODSS-19505 & 53.0166 \space\space -27.7448 & 3.59 & $1.40_{-1.25}^{+0.15}$ & $52.36_{-21.13}^{+23.19}$ & 1.4    \\ \hline
GOODSS-22610 & 53.0620 \space\space -27.7176 & 3.33 & $1.00_{-0.85}^{+0.50}$ & $11.29_{-4.72}^{+4.01}$ & 1.  0  \\ \hline
 \hline

UDS-1244 $\oplus$ & 34.2895 \space\space -5.2698 & 3.79 & $1.40_{-1.25}^{+0.05}$ & $137.60_{-72.10}^{+63.40}$ & 1.4    \\ \hline
UDS-2571 $\oplus$ & 34.2904 \space\space -5.2621 & 3.70 & $1.00_{-0.80}^{+0.50}$ & $69.39_{-26.50}^{+43.01}$ & 0.9    \\ \hline
UDS-7520 $\oplus$ & 34.2559 \space\space -5.2338 & 3.17 & $1.10_{-0.75}^{+0.75}$ & $274.00_{-149.50}^{+46.60}$ & 0.8    \\ \hline
UDS-10086 $\oplus$ & 34.3179 \space\space -5.2192 & 3.09 & $0.90_{-0.75}^{+0.95}$ & $31.87_{-15.19}^{+20.59}$ & 0.8    \\ \hline
UDS-10430 $\oplus$ & 34.2806 \space\space -5.2172 & 4.13 & $0.70_{-0.50}^{+0.60}$ & $105.30_{-56.97}^{+79.60}$ & 0.6    \\ \hline
UDS-20843 $\oplus$ & 34.4961 \space\space -5.1610 & 3.73 & $1.40_{-1.15}^{+0.05}$ & $95.64_{-26.41}^{+53.56}$ & 1.3    \\ \hline
UDS-23628 $\oplus$ & 34.2426 \space\space -5.1431 & 4.25 & $1.10_{-0.90}^{+0.15}$ & $83.09_{-32.42}^{+76.71}$ & 1.1    \\ \hline
UDS-25688 $\oplus$ & 34.5266 \space\space -5.1360 & 3.08 & $1.70_{-1.20}^{+0.20}$ & $261.70_{-56.70}^{+50.40}$ & 1.4    \\ \hline
UDS-25893 $\oplus$ & 34.3996 \space\space -5.1363 & 4.49 & $1.10_{-0.75}^{+0.05}$ & $233.80_{-117.10}^{+173.50}$ & 0.7    \\ \hline
UDS-4332 & 34.4657 \space\space -5.2519 & 3.18 & $1.80_{-1.79}^{+0.00}$ & $178.50_{-128.59}^{+98.60}$ & 1.5    \\ \hline
UDS-7779 & 34.2589 \space\space -5.2323 & 3.14 & $0.25_{-0.00}^{+1.60}$ & $101.30_{-35.36}^{+66.10}$ & 0.1    \\ \hline
UDS-8682 & 34.2937 \space\space -5.2270 & 3.46 & $1.55_{-1.40}^{+0.10}$ & $73.66_{-38.80}^{+49.94}$ & 1.5    \\ \hline
UDS-8689 & 34.2741 \space\space -5.2274 & 3.22 & $1.40_{-1.15}^{+0.25}$ & $175.50_{-79.74}^{+59.90}$ & 1.4    \\ \hline
UDS-11532 & 34.4207 \space\space -5.2116 & 4.21 & $0.70_{-0.55}^{+0.55}$ & $17.76_{-7.94}^{+13.04}$ & 0.7    \\ \hline
UDS-12640 & 34.5341 \space\space -5.2050 & 3.61 & $0.30_{-0.15}^{+1.25}$ & $45.98_{-14.58}^{+39.41}$ & 0.2    \\ \hline
UDS-32406 & 34.5426 \space\space -5.1861 & 3.28 & $0.50_{-0.35}^{+1.25}$ & $18.67_{-9.51}^{+9.99}$ & 0.3    \\ \hline

\caption{Physical properties of the red and dead candidates belonging to the ``reference'' sample, as obtained from their best fit with the top-hat SFH library without emission lines. ID$_{\textrm{CAND}}$ is the identification number in the CANDELS catalogues, followed by the WCS RA-DEC coordinates; $z_{\textrm{CAND}}$ is the official CANDELS redshift; Age is the time span from the onset of the star formation activity to the epoch of observation; $M_*$ is the stellar mass, which assumes a Salpeter IMF; $\Delta t_{\textrm{burst}}$ is the duration of the star formation activity, during which the SFR is constant ($SFR = M*/\Delta t_{\textrm{burst}}$). The SFR at the epoch of observation is always zero, by definition.
For each field, the table lists first the candidates which have also passed 
the ``lines'' selection (marked with a $\oplus$), and then the remaining objects in the ``reference'' sample. COSMOS ID-16676 is marked with a * because its properties have been obtained fitting at $z_{\textrm{spec}}=3.72$ rather than at $z_{\textrm{CANDELS}}=4.13$, which would yield: Age$=0.65_{-0.35}^{+0.65}$, $M_*=362.20_{-178.10}^{+231.10}$, $\Delta t_{\textrm{burst}} = 0.3$.}\label{tab1}
\end{longtable}

\newpage

\section{Examples of snapshots and SEDs of red and dead candidates} \label{examples}

\centering
\includegraphics[width=18cm]{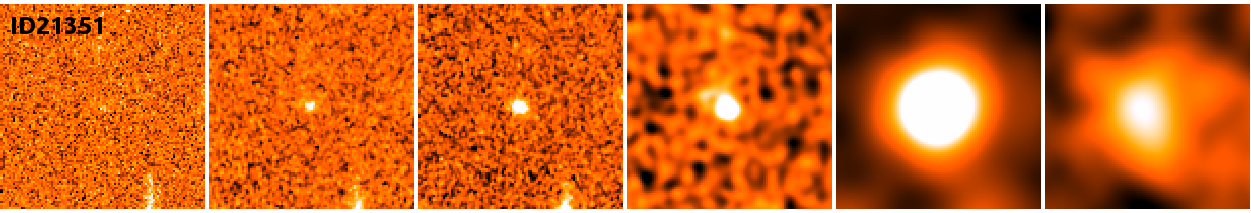}
\includegraphics[width=18cm]{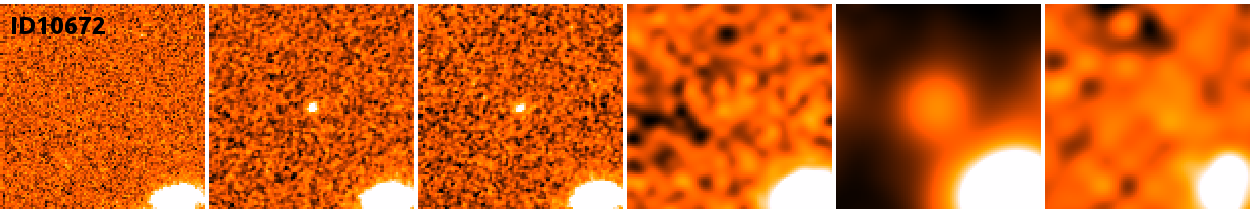}
\includegraphics[width=18cm]{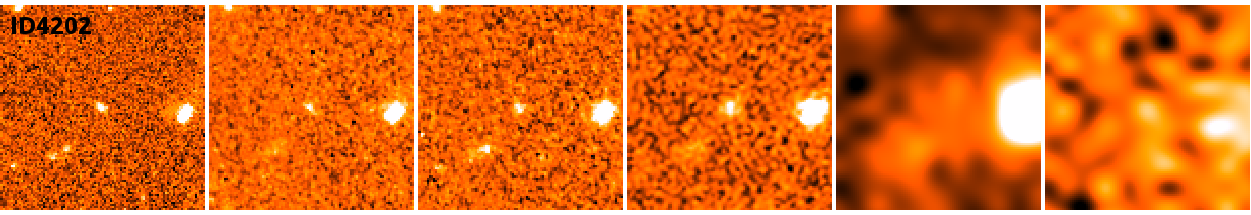}
\includegraphics[width=18cm]{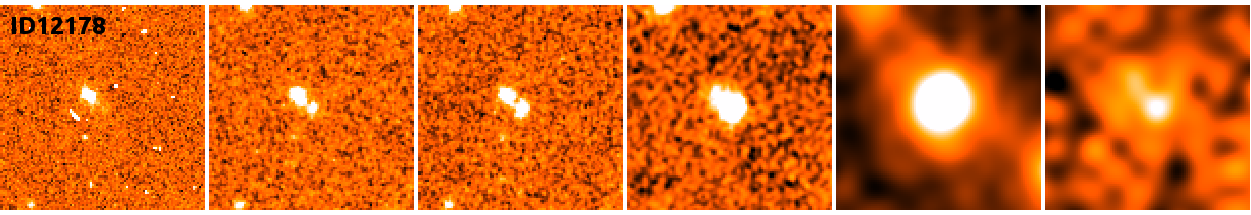}
\captionof{figure}{Illustrative snapshots of some passive candidates. Left to right: $ACS$ $ B435+V606+I814$ stack, $WFC3$ $J125$, $WFC3$ $H160$, $Ks$, IRAC $3.6 + 4.5$ $\mu$m stack, IRAC $5.8 + 8.0$ $\mu$m stack. Top to bottom: EGS-21351 (a robust candidate from the ``lines'' selection); GOODSN-10672 (the $z_{\textrm{CANDELS}}=6.7$ candidate discussed in Sect. \ref{z7}, again in the ``lines'' selection); GOODSS-4202 (a candidate from the ``reference'' selection, with a star-forming best fit in the run including emission lines; see also Fig. \ref{seds}); and GOODSS-12178 (the candidate from the ``lines'' selection which has VANDELS $z_{\textrm{spec}}=0.56$, but has a close companion which might contaminate the spectrum, as discussed in Sect. \ref{zspec}; this object has a passive best-fit in both the runs with and without emission lines, but it is excluded from the ``lines'' selection because it also has star-forming solutions with $p_{\textrm{SF}}>5\%$).}

\newpage
\centering
\includegraphics[width=7cm]{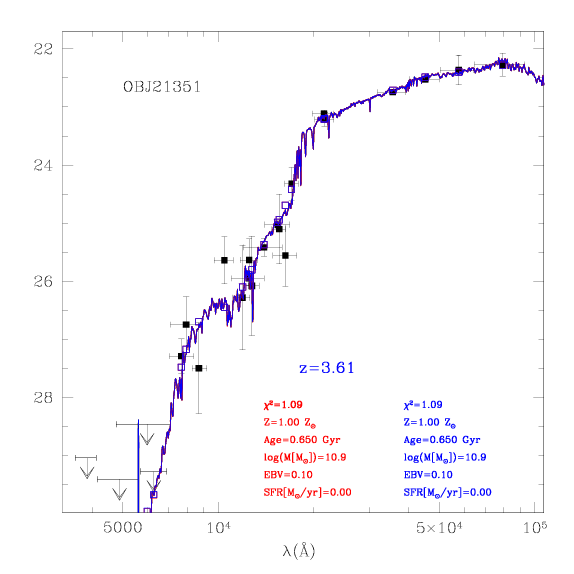}
\includegraphics[width=7cm]{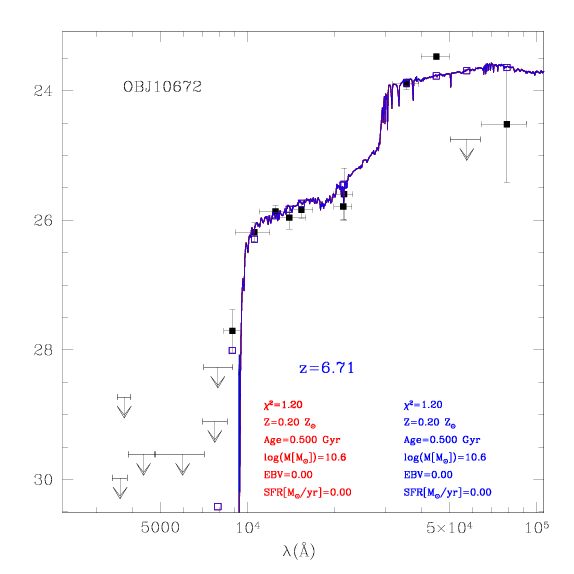}
\includegraphics[width=7cm]{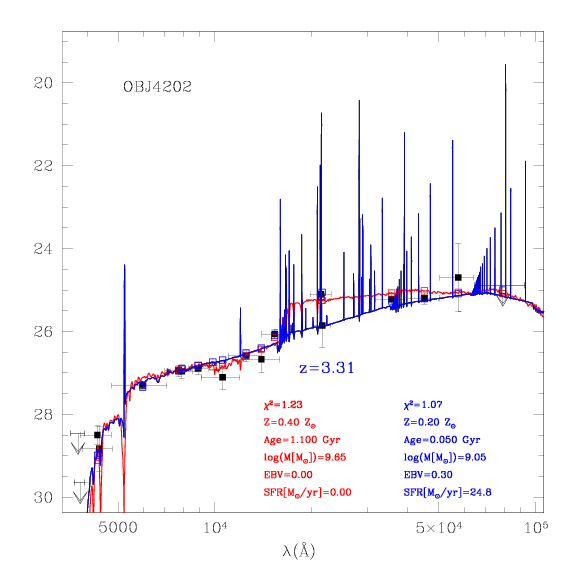}
\includegraphics[width=7cm]{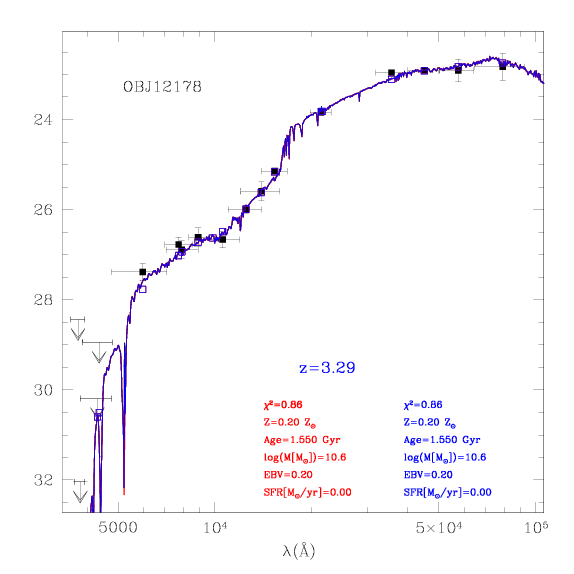}
\captionof{figure}{SEDs of the four passive candidates shown above. Black filled squares with error bars are the observed magnitudes. The red line is the SED of the best fit without emission lines, whereas the blue one corresponds the fit including the lines; empty squares of the same colors are the model magnitudes corresponding to the observed ones.} \label{seds}


\bsp	
\label{lastpage}
\end{document}